\documentclass[letterpaper,12pt]{article}   

\usepackage{osajnl2} 

\usepackage{bm}
\usepackage{amssymb}
\usepackage{amsmath}

\begin{document}

\newcommand{\modeprop}[0]{\mu}
\newcommand{\pwprop}[0]{\gamma}


\title{Modal formulation for diffraction by absorbing photonic crystal slabs}

\author{Kokou B. Dossou$^{1,*}$, Lindsay C. Botten$^1$, Ara A. Asatryan$^1$, \\
Bj\"{o}rn C.P. Sturmberg$^2$, Michael A. Byrne$^1$,  Christopher G. Poulton$^1$, \\
Ross C. McPhedran$^2$, and  C. Martijn de Sterke$^2$}

\address{$^1$ CUDOS, University of Technology, Sydney, N.S.W. 2007, Australia}

\address{$^2$CUDOS and IPOS, School of Physics, University of Sydney, NSW 2006, Australia}

\address{$^*$Corresponding author: Kokou.Dossou@uts.edu.au}


\begin{abstract}
A finite element-based modal formulation of diffraction of a plane wave by an absorbing photonic crystal slab
of arbitrary geometry is developed for photovoltaic applications.
The semi-analytic approach allows efficient and accurate calculation of the absorption of an array with a complex unit cell.
This approach gives direct physical insight into the absorption mechanism in such structures, which can be used to enhance the absorption.
The verification and validation of this approach is applied to a silicon nanowire array and the efficiency and accuracy of the method is demonstrated.
The method is ideally suited to studying the manner in which spectral properties (e.g., absorption)
vary with the thickness of the array, and we demonstrate this with efficient calculations which can identify an
optimal geometry.
\end{abstract}



\maketitle

%
%

\section{Introduction}
\label{intro}

Photonic crystals, which consist of a periodically arranged lattice of dielectric scatterers,
 have attracted substantial research interest over the past decade \cite{Joan}.
 Most commonly these structures are used to trap, guide, and otherwise manipulate pulses of light,
and have led to a variety of important applications in modern nanophotonics, including extremely
 high-Q electromagnetic cavities \cite{Noda} and slow-light propagation of electromagnetic
  pulses \cite{Krauss}.
However, a new and increasingly important application of photonic crystal structures
 is in the field of photovoltaics.
It has long been known that structured materials can be used to achieve photovoltaic
conversion efficiency beyond the Yablonovitch limit \cite{Fan:OE:2010,Yablonovitch:1982},
and indeed researchers have recently proposed photonic crystals \cite{PCcell,Chut}
and arrays of dielectric nanowires \cite{Tian,LinPov} as inexpensive ways to create
highly efficient absorbers.
Recent research in this area has been driven by advances in nanostructure fabrication
\cite{Lew} concurrent with increased investment in renewable energy technology
(for the latest developments see~\cite{Wur}).

One aspect in which the use of photonic crystals in photovoltaics differs markedly from
their employment in optical nanophotonics
is the important role played by material absorption. For most nanophotonic applications
absorption is an undesirable effect which must in general be minimized and can in many cases
be neglected. This is in stark
 contrast to photovoltaics, in which the main aim is to
exploit the properties of the structure to increase the overall absorption efficiency.

Modeling of absorbing photonic crystals has thus far been performed using direct numerical
methods such as the Finite-Difference Time-Domain (FDTD) method \cite{Tian},
the Finite Element Method (FEM) \cite{Li} or using the Transfer Matrix method \cite{LinPov}.
Although these methods have produced valuable information about the absorption
properties of such structures they do not allow us to gain direct physical
insight into the mechanism of the absorption within them.
Furthermore, these calculations require substantial computational time and resources.

Here we present a rigorous modal formulation of the scattering and absorption
of a plane wave by an array of absorbing nanowires, or, correspondingly, an absorbing photonic
 crystal slab (Fig.~\ref{slab}).
The approach is a generalization of diffraction by capacitive grids formulated
initially in Refs \cite{jewa1,jewa2} for perfectly conducting cylinders.
In contrast to conventional photonic crystal calculations the material absorption
is taken into account rigorously, using measured values for the real and the
imaginary parts of the refractive indices of the materials comprising the array.
Our formulation can be applied to array elements of arbitrary composition and cross-section.
The semi-analytical nature of this modal approach results in a method which
is quick, accurate, and gives extensive physical insight into the importance of the various
absorption and scattering mechanisms involved in these structures.

The method is based on an expansion in terms of the fundamental eigenmodes of the
structure in each different region. Within the
photonic crystal slab the fields are expanded in terms of Bloch modes, while the
fields in free space above and below the slab are expanded in a basis of plane waves.
These expansions are then matched using the continuity of the tangential components of
the fields at the top and the bottom interfaces of the slab to compute Fresnel interface reflection and
transmission matrices.

An important aspect of this approach is that
the set of Bloch modes must form a complete basis. This is not a trivial matter, as even for
structures consisting of lossless materials the eigenvalue problem for the Bloch modes is not
formally Hermitian.
However, it is well known from the classical treatment of non-Hermitian eigenvalue problems
\cite[p. 884]{Morse&Feshbach} that a complete basis may be obtained by including the Bloch modes of
the adjoint problem.
Here, we use the FEM to compute both the Bloch modes and the adjoint
 Bloch modes. Because the mode computation may be carried out in two dimensions (2D)
like waveguide mode calculations,
this routine is highly  efficient. We note that a previous study \cite{Blad:OE:2009},
undertaken using a different method,
failed to locate some of the modes, as is demonstrated in Sec.~\ref{Section:Dispersion};
we emphasis that the algorithm presented here is capable of generating a complete set of modes.
In addition the FEM allows us to compute Bloch modes for arbitrary materials
 and cross sections.

We note that this approach differs from earlier formulations \cite{jewa1,jewa2}
 in which the photonic crystal had to consist of an array of cylinders with a circular cross-section.
 With the modes identified, the transmission through, and reflection from, the slab can be computed
 using a generalization of Fresnel reflection and transmission matrices, and the absorption is
  found using an energy conservation relation.

The organization of the paper is as follows. The theoretical foundation of the method
is given in Sec.~\ref{formulation} while  the numerical verification, validation and
characterization of the absorption properties of a particular silicon nanowire array
occurs in Sec.~\ref{Applica}. The details of the mode orthogonality, normalization,
 completeness as well as energy and reciprocity relations are given in the appendices.

%
%

\section{Theoretical description}
\label{formulation}

As mentioned in Sec.~\ref{intro}, we separate the solution of the
diffraction problem into three steps, one involving the
consideration of the scattering of plane waves at the top interface and
the introduction of the Fresnel reflection and the transmission
matrices for a top interface (Fig.~\ref{slab}). Next we introduce the Fresnel
reflection and transmission matrices for the bottom interface by
considering the reflection of the waveguide modes of the semi-infinite array
of cylinders at the bottom interface.
Then the total reflection and transmission through the slab can be calculated
using a Fabry-Perot style of analysis.
The approach is based   on the calculation of the Bloch modes
and adjoint Bloch modes of an infinite array of cylinders.
Before doing so however, we first provide the field's
plane wave expansions above and below the photonic crystal slab.

%
%

\subsection{Plane wave expansion}
\label{PW}

\begin{figure}
\centerline{
\includegraphics[width=5.5cm]{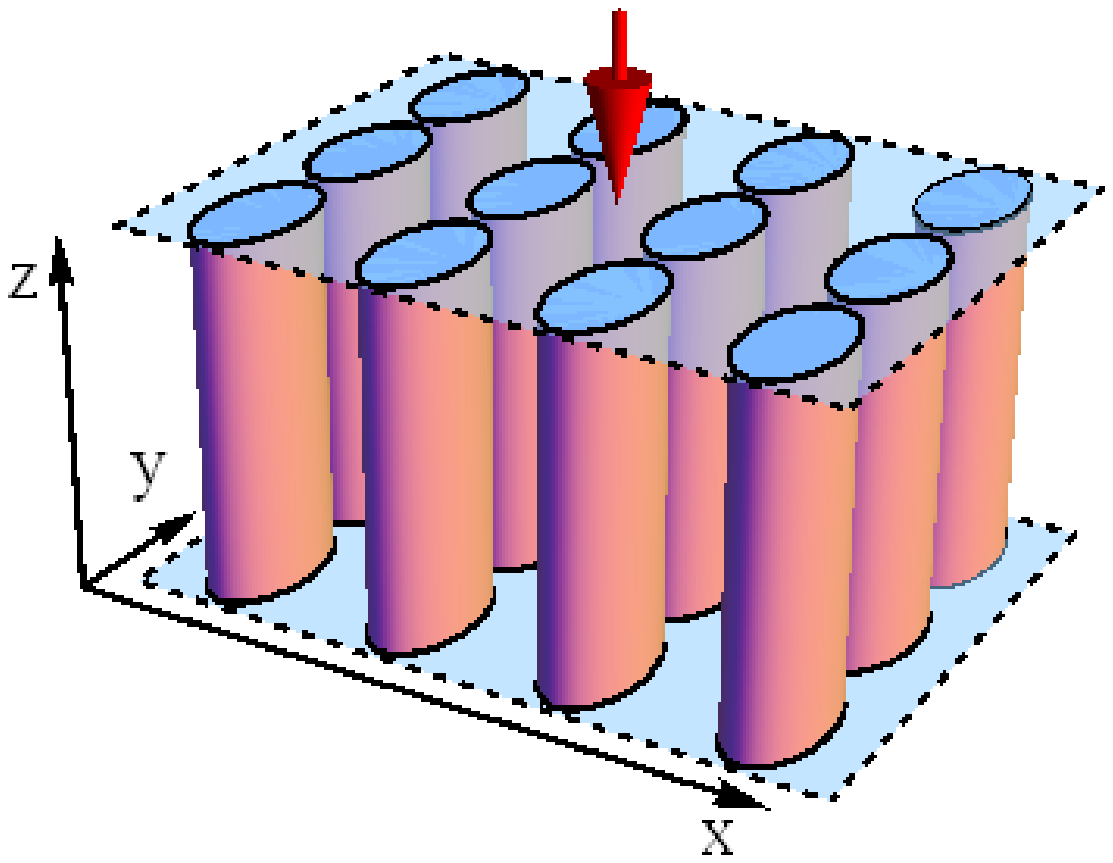}
\includegraphics[width=2cm]{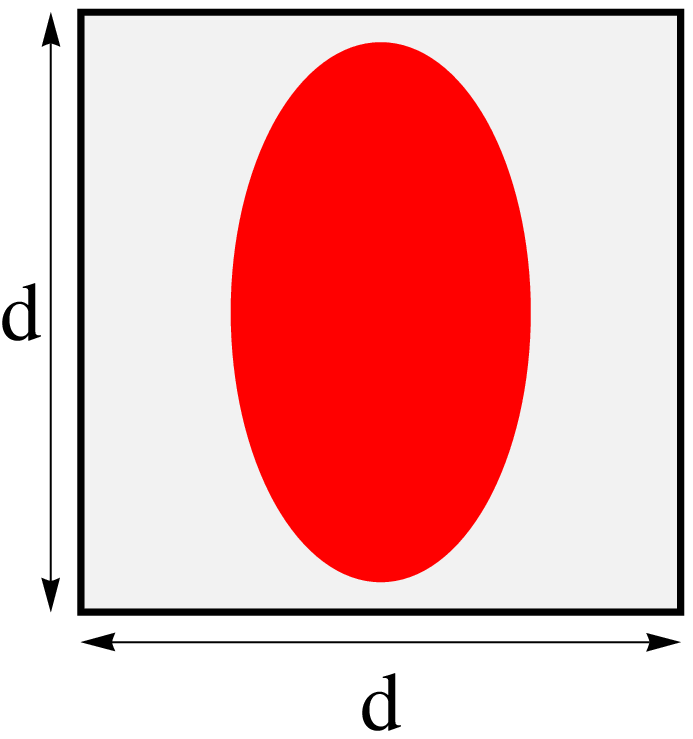}
}
\caption{Geometry of the problem. The dielectric
permittivity and the magnetic permeability of the slab can be
dispersive and can have absorption. The unit cell of the array
(right panel) can have arbitrary number of inclusions with arbitrary
shapes. The direction of the plane wave incident from above the
structure can be arbitrarily chosen.}
\label{slab}
\end{figure}

In a uniform media such as free space, all components of the electromagnetic field of a plane wave
must satisfy the Helmholtz equation
\begin{equation}
\label{waveHelmFS}
\nabla^2 \bm{E} + k^2 \bm{E} = 0,
\end{equation}
where  $k=2\pi/\lambda$ is the free space wavenumber.
Here we  consider the diffraction of a plane wave on a periodic square array of cylinders with finite
length
(see Fig.~\ref{slab}).
In such a structure, the fields have a
quasi-periodicity imposed by the incident plane wave field
$\exp (i(\alpha_0 x + \beta_0 y-\gamma_{00}z))$, where $\gamma_{00}=\sqrt{k^2-\alpha_0^2-\beta_0^2}$. That is,
\begin{equation}
\label{QP}
\bm{E}(\bm{r}+\bm{R})=\bm{E}(\bm{r})e^{i\bm{k}_{0}\cdot \bm{R}},
\end{equation}
where $\bm{k}_0=(\alpha_0, \beta_0)$ and $\bm{R}=(s_1 d,s_2 d)$ is a lattice vector, where $s_1$ and $s_2$
are integers. All plane waves of the form
$\exp (i(\alpha x + \beta y\pm\gamma z))=\exp(i(\bm{k}_{\perp}\cdot \bm{r}))
\exp(\pm i
{\mbox{\ensuremath $\gamma$}}
z)$,
 must satisfy the Bloch condition
{Eq.}~(\ref{QP})
and so
$e^{i\bm{k}_{\perp}\cdot \bm{R}_s}=e^{i \bm{k}_0 \cdot \bm{R}_s}$.
Hence $(\bm{k}_{\perp}-\bm{k}_0)\cdot \bm{R}_s=2\pi m$ , where $m$ is an integer. It follows then
that the coefficients $\alpha$ and $\beta$ are discretized as follows
\begin{eqnarray}
\alpha_p & = & \alpha_0 + p\,\frac{2 \pi}{d} \,,\\
\beta_q & = & \beta_0 + q\,\frac{2 \pi}{d} \,
\end{eqnarray}
and form the well known  diffraction grating orders.

We split the electromagnetic field into its transverse electric
 $TE$ and the transverse magnetic $TM$ components (see, for example, Ref.
\cite{Kan}) . For the transverse electric mode,
the electric field is perpendicular to the plane of incidence,
 while for the transverse magnetic mode the
magnetic field vector is perpendicular to the plane of incidence---with the plane of incidence
being defined by the $z$-axis and the plane wave propagation direction given by the vector $\bm{k}_0$.
These TE and TM resolutes are given by
\begin{eqnarray}
 \bm{R}^E_s (x,y) &=& -\frac{i}{Q_s} \bm{e}_z \times \nabla_\perp V_s  \nonumber\\
 &=&\frac{\bm{e}_z \times \bm{Q}_s}{Q_s} V_s(x,y),
 \label{RSE}\\
\bm{R}^M_s (x,y) &=& -\frac{i}{Q_s}  \nabla_\perp V_s  =
 \frac{\bm{Q}_s}{Q_s} V_s(x,y),
 \label{RSM}\\
 \bm{ Q}_s & = & (\alpha_p,\beta_q),
\end{eqnarray}
respectively,
where
$V_s(x,y) = {\exp(i \, \bm{Q}_s \cdot \bm{r})}$.
The TE and TM plan wave modes are mutually orthogonal, and are normalized such that
\begin{eqnarray}
 \iint \bm{R}^a_p \cdot \overline{\bm{R}}^b_q \,\, dS &=& \delta_{ab}
 \delta_{pq},
 \label{orRERM}
\end{eqnarray}
where the overline in $\overline{\bm{R}}^b_q$ denotes complex
conjugation and the integration is over the unit cell.


The general form of the plane wave expansions above and below
the grating (see Fig.~\ref{slab})
can then be written in terms of these $TE$ and $TM$ modes.
Following the nomenclature of Ref. \cite{Kan}, the plane wave expansions take the forms
\begin{eqnarray}
\bm{E}_\perp(\bm{r}) & = &  \sum_{s}
\chi_s^{-1/2}
\left[ f^{E-}_s e^{-i \pwprop_s(z-z_0)}\right. \nonumber\\
&+& \left. f^{E+}_s e^{i \pwprop_s(z-z_0)} \right] \bm{R}^E_s(\bm{r}) \nonumber\\
& + & \chi_s^{1/2}
\left[ f^{M-}_s e^{-i \pwprop_s(z-z_0)} \right.\nonumber\\
&+& \left. f^{M+}_s e^{i \pwprop_s(z-z_0)} \right] \bm{R}^M_s(\bm{r})
\label{epw}\\
\bm{e}_z\times \bm{H}_\perp(\bm{r}) & = &  \sum_{s}
\chi_s^{1/2}
\left[ f^{E-}_s e^{-i \pwprop_s(z-z_0)} \right.\nonumber\\
&-& \left. f^{E+}_s e^{i \pwprop_s(z-z_0)} \right] \bm{R}^E_s(\bm{r}) \nonumber\\
& +& \chi_s^{-1/2} \left[ f^{M-}_s e^{-i \pwprop_s(z-z_0)} \right.\nonumber\\
&-&\left.
f^{M+}_s e^{i \pwprop_s(z-z_0)} \right] \bm{R}^M_s(\bm{r})
\label{kpw}
\label{pwexpansion}
\end{eqnarray}
where $f^{E\pm}_s$ and $f^{M\pm}_s$ represent the amplitudes of transverse
 electric and magnetic component of the downward $(-)$ and upward $(+)$
 propagating plane waves and
\begin{eqnarray}
\label{eq1:PW:dispersion}
\pwprop_s & =& \sqrt{k^2-\alpha_p^2-\beta_q^2},
\end{eqnarray}
where $s$ denotes a plane wave channel represented by the pair of integers
 $s= (p, q) \in \mathbb{Z}^2$.
In the numerical implementation it
 is convenient to order the plane waves in descending order of $\pwprop_s^2$.
In Eqs (\ref{epw}) and (\ref{kpw}), $\chi_s$ is defined as
\begin{eqnarray}
\chi_s & = & \frac{\pwprop_s}{k},
\end{eqnarray}
and  the
factors $\chi_s^{\pm1/2}$  are included to normalize the calculation
of energy fluxes.

%

\subsection{FEM calculation of modes and adjoint modes of cylinder arrays}
\label{BM}

The FEM presented here is a general purpose numerical method which
can handle the square, hexagonal or any other array geometry. The
constitutive materials of the array can be dispersive and lossy. We
first introduce the eigenvalue problem and then we  present
a variational formulation and the corresponding FEM discretization.

%
%

\subsubsection{Maxwell's equations}

At a fixed frequency, the electric and magnetic fields of the
electromagnetic modes satisfy Maxwell's equations
\begin{eqnarray}
\nabla \times \bm{ H} =
& - & i\, k\, \varepsilon \,\bm{ E} \,, \quad \nabla \cdot ( \varepsilon \, \bm{E} ) = 0 \,,
\label{a1} \\
\nabla \times \bm{E} =
&   & i\, k\, \mu \, \bm{H} \,, \quad \nabla \cdot ( \mu \, \bm{H}) = 0 \, ,
\label{a2}
\end{eqnarray}
where $\varepsilon$ and $\mu$ are the relative dielectric permittivity and
magnetic permeability respectively. We  express the time dependence in the
form $e^{-i\omega t}$. The magnetic field $\bm{H}$ has been rescaled
as $Z_0 \bm{H}\rightarrow\bm{H}$ with
$Z_0=\sqrt{\mu_0/\varepsilon_0}$, the impedance of free space.
We now consider the electromagnetic modes of an array of cylinders
of infinite length. The cylinder axes are aligned with the $z-$axis.
The dielectric permittivity and magnetic permeability
of the array are invariant with respect to $z$. From this
translational invariance, we know that the Bloch modes of the array
have a $z-$dependence $\exp(i \zeta z)$ and are quasi-periodic with
respect to  $x$ and $y$. This reduces the problem of finding the
modes to a unit cell $\Omega$ in the $xy-$plane (see
Fig.~\ref{slab}).

As explained in Section~\ref{section:Adjoint}, this modal problem for the cylinder arrays
is not Hermitian and therefore the eigenmodes do not necessarily form an orthogonal set.
However, by introducing the modes of the adjoint problem we can form
a set of adjoint modes which have a biorthogonality
property~\cite{Botten:OA:1981} with respect to the primal
eigenmodes.
In order to introduce the adjoint problem, we first write Maxwell's equations for
conjugate material parameters $\overline{\varepsilon}(\bm{r})$ and $\overline{\mu}(\bm{r})$:
\begin{eqnarray}
\nabla \times \bm{H}^{\rm c} =
& - & i\, k\, \overline{\varepsilon} \, \bm{E}^{\rm c} \,, \quad
\nabla \cdot ( \overline{\varepsilon} \, \bm{E}^{\rm c}) = 0 \,,
 \\
\nabla \times \bm{E}^{\rm c} =
&   & i\, k\, \overline{\mu} \, \bm{H}^{\rm c} \,, \quad
\nabla \cdot ( \overline{\mu} \, \bm{H}^{\rm c} ) = 0 \, .
\end{eqnarray}
The fields $\bm{E}^{\rm c}$ and $\bm{H}^{\rm c}$ have the same time dependence and
quasi-periodicity as $\bm{E}$ and $\bm{H}$.
We define the adjoint modes as the conjugate fields
$\bm{E}^{\dagger} = \overline{\bm{E}^{\rm c}}$ and $\bm{H}^{\dagger}
= \overline{\bm{H}^{\rm c}}$, and they satisfy Maxwell's
equations:
\begin{eqnarray}
\nabla \times \bm{H}^{\dagger} =
&  & i\, k\, \varepsilon \, \bm{E}^{\dagger} \,, \quad
\nabla \cdot ( \varepsilon \, \bm{E}^{\dagger} ) = 0 \,,
\label{eq2:a1} \\
\nabla \times \bm{E}^{\dagger} =
&  - & i\, k\, \mu \, \bm{H}^{\dagger} \,, \quad
\nabla \cdot ( \mu \, \bm{H}^{\dagger} ) = 0 \, .
\label{eq2:a2}
\end{eqnarray}
Therefore, the adjoint modes $\bm{E}^{\dagger}$ and $\bm{H}^{\dagger}$ satisfy the same wave
equations as $\bm{E}$ and $\bm{H}$, but have
the opposite quasi-periodicity and $e^{i \, \omega \, t}$ time dependence.
%

%
%

\subsubsection{Variational formulation of the eigenvalue problem}
\label{VF}

Let $\Omega$ denote a unit cell of the periodic lattice.
Within the array, a Bloch mode is a nonzero solution of the vectorial wave equation
\begin{equation}
\label{wave1}
\nabla \times (\mu^{-1} \nabla \times \bm{E}) - k^2 \varepsilon \bm{E} = 0,
\; \mbox{on the domain $\Omega$,}
\end{equation}
which is quasi-periodic in the transverse plane with respect to the
wave vector $\bm{k}_{\perp}$, and has exponential $z$ dependence,
with the propagation constant $\zeta$, i.e.
\begin{equation}
\label{mode1}
\bm{E}(x,y,z) = \bm{E}(x,y) \, e^{i \, \zeta \, z}.
\end{equation}
The longitudinal and transverse components of the electric field $\bm{E}$ are respectively
$E_z$ and $\bm{E}_{\perp} = [ E_x, E_y]^T${\underline{..}}
At the edges of the unit cell, the tangential component
$\bm{E}_{\perp} \cdot \vec{\tau}$ ($\vec{\tau}$ denotes a unit
tangential vector to the unit cell boundary $\partial \Omega$)
and the longitudinal component $E_z$ of a Bloch mode satisfy the
boundary conditions
\begin{eqnarray}
\label{boundary}
\mbox{$\bm{E}_{\perp} \cdot \vec{\tau}$ and $E_z$ are
$\bm{k}_{\perp}$-quasi-periodic over $\partial \Omega$} .
\end{eqnarray}
The quasi-periodicity of the components $\bm{H}_{\perp} \cdot
\vec{\tau}$ and $H_z$ of the magnetic field $\bm{H}$ associated with
$\bm{E}$ is enforced as a ``natural boundary condition" of the FEM.

Taking the exponential $z$ dependence into account by substituting
(\ref{mode1}) into (\ref{wave1}), one is led to the coupled partial
differential equations
\begin{equation}
\label{mode2}
\left\{ \begin{array}{l}
\vec{\nabla}_{\perp} \times (\mu^{-1} (\nabla_{\perp} \times \bm{E}_{\perp})) -
i \, \zeta \mu^{-1} \nabla_{\perp} E_z
\\
\hspace*{3.2cm} + (\zeta^2 \mu^{-1} - k^2 \varepsilon) \bm{E}_{\perp} = 0 , \\
\!\! -i \, \zeta \nabla_{\perp}
\! \cdot \!
(\mu^{-1} \bm{E}_{\perp})
\!\!- \!
\nabla_{\perp}
\!\! \cdot \!
(\mu^{-1} \nabla_{\perp} E_z)
\! - \!
k^2 \varepsilon E_z
\! = \!
0
\end{array} \right.
\end{equation}
where the operators $\nabla_{\perp} E_z$ and $\nabla_{\perp} \cdot \bm{E}_{\perp}$
are the gradient and the divergence
with respect to the transverse variables $x$ and $y$; the transverse curl operators are
defined as
\begin{eqnarray}
\vec{\nabla}_{\perp} \times \bm{e}_zF_z & = &
\left[ \begin{array}{r} \frac{\partial F_z}{\partial y} \\[6pt]
{\mbox{\ensuremath $-$}}
\frac{\partial F_z}{\partial x} \end{array} \right],
\\
\nabla_{\perp} \times \bm{F}_{\perp}
& = &
\left(
\frac{\partial F_y}{\partial x} -
\frac{\partial F_x}{\partial y}
\right) \bm{e}_z.
\end{eqnarray}
Problem~(\ref{mode2}) is a nonlinear eigenproblem with respect to
the unknown $\zeta$ since it involves both $\zeta$ and $\zeta^2$.
For $\zeta \neq 0$, the substitution
\begin{equation}
\label{Ez:scaled}
E_z = -i \, \zeta \hat{E}_z
\end{equation}
leads to a generalized eigenvalue problem involving only~$\zeta^2$
\begin{equation}
\label{mode3}
\left\{ \begin{array}{l}
\vec{\nabla}_{\perp} \times (\mu^{-1} (\nabla_{\perp} \times \bm{E}_{\perp}) )
- k^2 \varepsilon \bm{E}_{\perp} \\
\hspace*{2.5cm} = \zeta^2( \mu^{-1} \nabla_{\perp} \hat{E}_z - \mu^{-1} \bm{E}_{\perp}), \\
\!\! - \nabla_{\perp}
\! \cdot \!
(\mu^{-1} \bm{E}_{\perp})
\! + \!
\nabla_{\perp}
\! \cdot \!
(\mu^{-1} \nabla_{\perp} \hat{E}_z)
+ k^2 \varepsilon \hat{E}_z
\! = \! 0.
\end{array} \right.
\end{equation}
Note that an eigenvalue $\zeta^2 \neq 0$ of Eq.~(\ref{mode3})
corresponds to a pair of propagation constants $\zeta^+ =
\zeta$ and $\zeta^- = -\zeta$ which are respectively associated with
a \emph{upward propagating} (i.e., towards $z \to \infty$) wave,
$\bm{E}^+ = [\bm{E}_{\perp}, E_z] = [\bm{E}_{\perp}, -i \, \zeta^+
\hat{E}_z] $ (according to the scaling~(\ref{Ez:scaled})) and a
\emph{downward propagating} wave $\bm{E}^- = [\bm{E}_{\perp}, E_z] =
[\bm{E}_{\perp}, -i \, \zeta^- \hat{E}_z] $.
As shown in Ref.~\cite{Vardapetyan:MC:2003}, except for a countable set of frequencies,
in general all eigenvalues of the problem~(\ref{mode3}) are nonzero.
We remark that, in some cases, the mathematical analysis of the eigenproblem can be simpler and more elegant if
Eq.~(\ref{mode3}) is rewritten in the following form
\begin{equation}
\label{mode3b}
\left\{ \begin{array}{l}
\vec{\nabla}_{\perp} \times (\mu^{-1} (\nabla_{\perp} \times \bm{E}_{\perp}) )
- k^2 \varepsilon \bm{E}_{\perp} \\
\hspace*{2.5cm} = \zeta^2 \left( \mu^{-1} \nabla_{\perp} \hat{E}_z - \mu^{-1} \bm{E}_{\perp} \right), \\
0 \! = \! \zeta^2 \left(
\!\! - \nabla_{\perp}
\! \cdot \!
(\mu^{-1} \bm{E}_{\perp})
\! + \!
\nabla_{\perp}
\! \cdot \!
(\mu^{-1} \nabla_{\perp} \hat{E}_z)
+ k^2 \varepsilon \hat{E}_z \right).
\end{array} \right.
\end{equation}
For instance, as is explained below (see Eqs.~(\ref{primal:mode:1})--(\ref{operator:M})),
the differential operators on the left and right hand sides of Eq.~(\ref{mode3b})
are Hermitian, in the case of lossless gratings while, for lossy gratings, the adjoint of each operator
is the complex conjugate of the operator.
However, all nonzero fields of the form $\bm{E} = (0,0,\hat{E}_z)$ become eigenmodes
(although most are non-physical) of  Eq.~(\ref{mode3b}) associated with
the zero eigenvalue. In order to avoid the unnecessary calculations of these non-physical modes,
we have used  a numerical implementation based on Eq.~(\ref{mode3}).

We now explain our convention used in the modal classification.
Taking into account the exponential $z$ dependence $e^{i \, \zeta \,
z}$, if ${\rm Re} \, \zeta^2>0$ and ${\rm Im} \, \zeta^2 = 0$
(propagating mode) the upward travelling wave $\bm{E}^+$ corresponds
to the positive square root of $\zeta^2$ , i.e., $\zeta^+>0$;
otherwise
if ${\rm Re} \, \zeta^2<0$ or ${\rm Im} \, \zeta^2 \neq 0$
(evanescent mode) the upward travelling wave $\bm{E}^+$ corresponds
to the mode such that $|e^{i \, \zeta \, z}|$ decreases as $z$ increases, i.e.,
$\zeta^+$ is the complex square root of $\zeta^2$ such that ${\rm Im} \, \zeta^+ > 0$.
In order to obtain the variational formulation corresponding to the
problem
{Eqs.}~(\ref{mode3}) and (\ref{boundary})
we introduce the following
functional spaces
\begin{eqnarray}
\mathcal{V} & = &
\! \left\{\begin{array}{l}
 F_z \in L^2(\Omega) \; | \; \nabla_{\perp} F_z \in L^2(\Omega); \; \\
\hspace*{0.4cm} \mbox{$F_z$ is $\bm{k}_\perp$-quasi-periodic over $\partial \Omega$}
\end{array} \right\} , \\
\mathcal{W} & = &
\!
\left\{\begin{array}{l} \!\!
\bm{F}_{\perp} \in (L^2(\Omega))^2 \; | \; \nabla_{\perp} \times \bm{F}_{\perp} \in L^2(\Omega); \; \\
\!\!
\mbox{$\bm{F}_{\perp} \cdot \vec{\tau}$ is $\bm{k}_{\perp}$-quasi-periodic over $\partial \Omega$}
\end{array}
\! \right\} .
\end{eqnarray}
Then if we multiply the first and second equations in Eq.~(\ref{mode3}) respectively
by the complex conjugate of the test functions $\bm{F}_{\perp} \in \mathcal{W}$ and
$F_z \in \mathcal{V}$,
we obtain the variational formulation of the problem after integration by parts:

{\em Find $\zeta \in \mathbb{C}$ and $(\bm{E}_{\perp}, \hat{E}_z) \in \mathcal{W} \times \mathcal{V}$ such that
$(\bm{E}_{\perp}, \hat{E}_z) \neq 0$ and
$\forall (\bm{F}_{\perp}, F_z) \in \mathcal{W} \times \mathcal{V}$}
\begin{equation}
\label{formul1}
\left\{ \begin{array}{l}
\left( (\nabla_{\perp} \times \bm{F}_{\perp}) , \mu^{-1} (\nabla_{\perp} \times \bm{E}_{\perp}) \right)
- k^2 \left( \bm{F}_{\perp} , \varepsilon \bm{E}_{\perp} \right)\\
\hspace*{2cm} = \zeta^2 \left( \bm{F}_{\perp} , \mu^{-1} (\nabla_{\perp} \hat{E}_z - \bm{E}_{\perp}) \right), \\[6pt]
\left( \nabla_{\perp} F_z , \mu^{-1} \bm{E}_{\perp} \right)
- \left( \nabla_{\perp} F_z , \mu^{-1} \nabla_{\perp} \hat{E}_z \right) \\
\hspace*{2cm} + \, k^2 \left( F_z , \varepsilon \hat{E}_z \right)  = 0,
\end{array} \right.
\end{equation}
where $(,)$ represents the $L^2(\Omega)$ inner product
\begin{eqnarray}
\label{eq:inner:product}
(\bm{F}, \bm{E})
& = &
\int_{\Omega}
\overline{\bm{F}} \cdot \bm{E} \, dA \, .
\end{eqnarray}

%
%

\subsubsection{Finite element discretization}

Let $\mathcal{V}_h$ and $\mathcal{W}_h$ be two finite dimensional approximation spaces to
the functional spaces $\mathcal{V}$ and $\mathcal{W}$ respectively. The discretized
problem is obtained by substituting $\mathcal{W}_h \times \mathcal{V}_h$ for $\mathcal{W} \times \mathcal{V}$
in the formulation of problem~(\ref{formul1}). We
introduce sets of basis functions, respectively for the spaces $\mathcal{V}_h$ and $\mathcal{W}_h$, and map these onto
Eq.~(\ref{formul1}) to derive the
discretized problem in matrix form~\cite{Dossou:CMAME:2005}:
\begin{equation}
\label{matrix1}
\left[ \begin{array}{cc}
\bm{K}_{tt} & \bm{0} \\
\bm{K}_{zt} & \bm{K}_{zz}
\end{array} \right]
\left[ \begin{array}{c} \bm{E}_{\perp,n} \\ \hat{\bm{E}}_{z,n} \end{array} \right]
= \zeta_n^2
\left[ \begin{array}{cc}
\bm{M}_{tt} & \bm{K}_{zt}^H \\
\bm{0} & \bm{0}
\end{array} \right]
\left[ \begin{array}{c} \bm{E}_{\perp,n} \\ \hat{\bm{E}}_{z,n} \end{array} \right],
\end{equation}
where the superscript $H$ denotes the Hermitian transpose (or conjugate transpose) operator.
Let $( \bm{G}_{\perp,i} )_{i \in \{1,2,\dots, \dim (\mathcal{W}_h) \}}$
and $(\hat{G}_{z,i} )_{i \in \{1,2,\dots, \dim (\mathcal{V}_h) \}}$
be the chosen basis functions for the spaces $\mathcal{W}_h$ and $\mathcal{V}_h$ respectively.
For $i,j \in \{1,2,\dots, \dim (\mathcal{W}_h) \}$, the elements of the matrices
$\bm{K}_{tt}$ and $\bm{M}_{tt}$ are defined as
\begin{eqnarray}
(\bm{K}_{tt})_{ij} & = &
\left( (\nabla_{\perp} \times \bm{G}_{\perp,i}) , \mu^{-1} (\nabla_{\perp} \times \bm{G}_{\perp,j}) \right)
- k^2 \left( \bm{G}_{\perp,i} , \varepsilon \bm{G}_{\perp,j} \right) \\
(\bm{M}_{tt})_{ij} & = &
- \left( \bm{G}_{\perp,i} , \mu^{-1} (\nabla_{\perp} \bm{G}_{\perp,j}) \right) ,
\end{eqnarray}
and, for $i \in \{1,2,\dots, \dim (\mathcal{V}_h) \}$ and
$j \in \{1,2,\dots, \dim (\mathcal{W}_h) \}$,
the matrix $\bm{K}_{zt}$ is defined as
\begin{eqnarray}
(\bm{K}_{zt})_{ij} & = &
\left( \nabla_{\perp} \hat{G}_{z,i} , \mu^{-1} \bm{G}_{\perp,j} \right) ,
\end{eqnarray}
while, for $i,j \in \{1,2,\dots, \dim (\mathcal{V}_h) \}$, the coefficients
of the matrix $\bm{K}_{zz}$ are
\begin{eqnarray}
(\bm{K}_{zz})_{ij} & = &
- \left( \nabla_{\perp} \hat{G}_{z,i} , \mu^{-1} \nabla_{\perp} \hat{G}_{z,j} \right)
+ \, k^2 \left( \hat{G}_{z,i} , \varepsilon \hat{G}_{z,j} \right) .
\end{eqnarray}
The generalized eigenvalue problem
{Eq.}~(\ref{matrix1})
can be solved efficiently using the eigensolver
for sparse matrices ARPACK~\cite{ARPACK:SIAM:1998}.
Once the array modes $\bm{E}_n$ are computed, we express a field $\bm{E}$ inside
the array by the modal expansion
\begin{eqnarray}
\label{array:mode:expansion}
\bm{E}(x,y,z) & = & \sum_{n}
c_n \, \bm{E}_n (x,y) \, \exp(i \, \zeta_n \, z)
\end{eqnarray}
where the index $n$ counts out all the upward and downward propagating modes.

In contrast, a formulation of Maxwell's equations which is based on $(E_z, H_z)$ fields
leads to a nonlinear eigenvalue problem~\cite{Blad:OE:2009} which becomes inefficient to solve
as the number of eigenvalues to be computed increases.
The ARPACK library can be used to compute a few selected
eigenvalues of  large sparse matrices, and a shift and invert
spectral transformation can be applied so that the numerical
solutions converge to eigenvalues located within a desired region of
the spectrum. Here, by considering the plane wave dispersion
relation
{Eq.}~(\ref{eq1:PW:dispersion}),
the target region includes the
propagation constant $\zeta$ near a reference value
$\zeta_{\rm ref} = \left( n_{\rm ref}^2 k^2 - \alpha_0^2 - \beta_0^2 \right)^{1/2}$
where the reference index $n_{\rm ref} \in \mathbb{R}$ is selected to be slightly higher
than the largest real part of the slab refractive indices.

We have chosen $\mathcal{V}_h$ as the space of two-dimensional vector fields whose components
are piecewise polynomials of degree $p$
while $\mathcal{W}_h$ consists of piecewise continuous polynomials of degree $p+1$ (in this paper $p=2$).
The vector fields of $\mathcal{V}_h$ must be tangentially continuous across the inter-element edges of the finite element triangulation
while their normal component is allowed to be discontinuous ({\em edge element}).

We now present some general principles which have guided our choice for the spaces $\mathcal{V}_h$ and $\mathcal{W}_h$.
It is known that, for a simply connected domain $\Omega$, the
gradient, curl and div operators form an {\em exact sequence}:
\begin{equation}
\label{eq1:diagram}
\hspace*{-0.4cm}
\begin{array}{ccccccc}
H(\mbox{Grad},\Omega) & \!\!
\stackrel{\nabla}{\rightarrow} \!\! &
H(\mbox{Curl}, \Omega) & \!\!
\stackrel{\nabla \times}{\rightarrow} \!\! &
H(\mbox{Div}, \Omega) & \!\!
\stackrel{\nabla \cdot}{\rightarrow} \!\! &
L(\Omega)
\end{array}
\!\!
\end{equation}
i.e., the range of each operator coincides with the kernel of the following one.
The derivation of this statement is based on the Poincar\'e lemma and for more details
see Ref.~\cite[p. 133]{Bossavit}, for instance.
For the sake of numerical stability~\cite{Boffi:Springer:2008},
FEM approximation spaces must be chosen such that the exact sequence is reproduced at
the discrete level.

In the context of waveguide mode theory, a scalar function takes the form
$\left( F_z(x,y) \, e^{i \, \zeta \, z} \right)$ and its gradient is
$\nabla \left(F_z(x,y) \, e^{i \, \zeta \, z} \right)
= \left[ \nabla_{\perp} F_z(x,y),  \, i \, \zeta \, F_z(x,y) \right]
 \, e^{i \, \zeta \, z}$.
If $F_z(x,y)$ is a piecewise continuous polynomial of degree $p+1$,
i.e., $F_z(x,y) \in \mathcal{W}_h$, then the two components of $\nabla_{\perp}
F_z(x,y)$ are piecewise polynomials of degree $p$, i.e.,
$\nabla_{\perp} F_z(x,y) \in \mathcal{V}_h$. We can then verify that, with our
choice of $\mathcal{V}_h$ and $\mathcal{W}_h$, the first exactness relation of
Eq.~(\ref{eq1:diagram}), i.e., $H(\mbox{Grad},\Omega)
\stackrel{\nabla}{\rightarrow} H(\mbox{Curl}, \Omega)$, is also
reproduced at the discrete level. We do not discuss the second
exactness relation since, here, we do not have to build an
approximation space for $H(\mbox{Div}, \Omega)$.
We recall that in this paper we set $p=2$, and
approximate the transverse and longitudinal components using
polynomials of degrees 2 and 3 respectively; the construction of
the basis functions for the FEM spaces
is described in Ref.~\cite{Dossou:CMAME:2005}.

%
%

\subsubsection{Adjoint modes and the biorthogonality property}
\label{section:Adjoint}
Modal orthogonality relations, or more correctly biorthogonality relations, in
the case of the problem considered here, are important in determining the field expansion
coefficients $c_n$ in Eq.~(\ref{array:mode:expansion}).
Although we may recast Eq.~(\ref{matrix1}) in a form in which
each matrix is Hermitian (for the lossless case):
\begin{equation}
\label{matrix:sym2}
\left[ \begin{array}{cc}
\bm{K}_{tt} & \bm{0} \\
\bm{0} & \bm{0}
\end{array} \right]
\left[ \begin{array}{c} \bm{E}_{\perp,n} \\ \hat{\bm{E}}_{z,n} \end{array} \right]
= \zeta_n^2
\left[ \begin{array}{cc}
\bm{M}_{tt} & \bm{K}_{zt}^H \\
\bm{K}_{zt} & \bm{K}_{zz}
\end{array} \right]
\left[ \begin{array}{c} \bm{E}_{\perp,n} \\ \hat{\bm{E}}_{z,n} \end{array} \right],
\end{equation}
this generalized eigenvalue problem $\bm{A} \, \bm{v} = \zeta^2 \, \bm{B} \, \bm{v}$
is not Hermitian in general,  since for two Hermitian matrices, $\bm{A}$ and $\bm{B}$, the
corresponding eigenproblem,
$\bm{B}^{-1} \, \bm{A} \, \bm{v} = \zeta^2 \, \bm{v}$
is not necessarily Hermitian since the
product of two Hermitian matrices is not, in general, Hermitian.
Accordingly,  the eigenmodes $\bm{E}_n$ do not necessarily form an orthogonal set.
It is also clear that the eigenproblem is not Hermitian in the presence of loss.

Therefore we introduce the adjoint problem that we solve to obtain a set of adjoint modes
which have a biorthogonality property~\cite{Botten:OA:1981} with respect to the
eigenmodes $\bm{E}_n$.
In order to define the adjoint operators, we first write Eq.~(\ref{mode3b}),
the alternate form of Eq.~(\ref{mode3}), as
\begin{eqnarray}
\label{primal:mode:1}
\mathcal{L} \, \bm{E}_n
& = &
\zeta_n^2 \mathcal{M} \, \bm{E}_n ,
\end{eqnarray}
where $\mathcal{L}$ and $\mathcal{M}$ are the differential operators defined by
\begin{eqnarray}
\label{operator:L}
\mathcal{L} \, \bm{E}
\! = \!\!
\left[ \begin{array}{cc}
\vec{\nabla}_{\perp} \times (\mu^{-1} (\nabla_{\perp} \times \bm{E}_{\perp}) )
- k^2 \varepsilon \bm{E}_{\perp}
& 0 \\
0 & 0
\end{array} \right] \! , \\
\label{operator:M}
\mathcal{M} \, \bm{E}
 \! = \!\!
\left[ \begin{array}{cc}
- \mu^{-1} \bm{E}_{\perp} & \mu^{-1} \nabla_{\perp} \hat{E}_z \\
\! - \nabla_{\perp}
\! \cdot \!
(\mu^{-1} \bm{E}_{\perp} \!) &
\nabla_{\perp}
\! \cdot \!
(\mu^{-1} \nabla_{\perp} \hat{E}_z)
\! + \!
k^2 \varepsilon \hat{E}_z
\end{array}
\! \right] \! .
\end{eqnarray}

The adjoint operators $\mathcal{L}^{\dagger}$ and
$\mathcal{M}^{\dagger}$,
with respect to the inner product
{Eq.}~(\ref{eq:inner:product})
are
\begin{eqnarray}
\mathcal{L}^{\dagger} \bm{F}
 \! = \!\!
\left[ \begin{array}{cc}
\vec{\nabla}_{\perp} \times (\overline{\mu}^{-1} (\nabla_{\perp} \times \bm{F}_{\perp}) )
- k^2 \overline{\varepsilon} \bm{F}_{\perp}
& 0 \\
0 & 0
\end{array} \right] \! , \\
\mathcal{M}^{\dagger} \bm{F}
 \! = \!\!
\left[ \begin{array}{cc}
- \overline{\mu}^{-1} \bm{F}_{\perp} & \overline{\mu}^{-1} \nabla_{\perp} \hat{F}_z \\
\! - \nabla_{\perp}
\! \cdot \!
(\overline{\mu}^{-1} \bm{F}_{\perp}) &
\nabla_{\perp}
\! \cdot \!
(\overline{\mu}^{-1} \nabla_{\perp} \hat{F}_z)
\! + \!
k^2 \overline{\varepsilon} \hat{F}_z
\end{array}
\! \right] \!\! ,
\end{eqnarray}
and follow from the definitions
\begin{eqnarray}
\label{def:adjoint:L}
\left( \mathcal{L}^{\dagger} \bm{F} , \bm{E} \right)
& = &
\left( \bm{F} , \mathcal{L} \, \bm{E} \right),
\quad \forall  \bm{E} , \bm{F} ,
\\
\label{def:adjoint:M}
\left( \mathcal{M}^{\dagger} \bm{F} , \bm{E} \right)
& = &
\left( \bm{F} , \mathcal{M} \, \bm{E} \right),
\quad \forall  \bm{E} , \bm{F}  .
\end{eqnarray}
Although for lossless media, we then have $\mathcal{L}^{\dagger} =
\mathcal{L}$ and $\mathcal{M}^{\dagger} = \mathcal{M}$, as in
Eq.~(\ref{matrix:sym2}), eigenproblem Eq.~(\ref{primal:mode:1}) is
not Hermitian
and, as we see in Section~\ref{Applica}, complex values of
$\zeta_n^2$ can occur even for a lossless photonic crystal.

The modes $\bm{E}^c_n$ are the eigenmodes of the problem
\begin{eqnarray}
\label{adjoint:mode:1}
\mathcal{L}^{\dagger} \bm{E}^c_m
& = &
{\zeta^c_m}^2 \mathcal{M}^{\dagger} \bm{E}^c_m
\end{eqnarray}
which satisfy the same quasi-periodic boundary conditions as $\bm{E}_n$.
We now conjugate the boundary value problem
{Eq.}~(\ref{adjoint:mode:1})
and
take into account the fact that
$\mathcal{L}^{\dagger} = \overline{\mathcal{L}}$
and
$\mathcal{M}^{\dagger} = \overline{\mathcal{M}}$.
Consistent with Eqs~(\ref{eq2:a1}) and (\ref{eq2:a2})
we redefine the adjoint mode as the eigenmode
$\bm{E}^{\dagger}_m$
which satisfies the same partial differential equation as
$\bm{E}_n$, i.e.,
\begin{eqnarray}
\label{adjoint:mode:2}
\mathcal{L} \, \bm{E}^{\dagger}_m
& = &
{\zeta^{\dagger}}_m^2 \mathcal{M} \, \bm{E}^{\dagger}_m,
\end{eqnarray}
but with the quasi-periodic boundary conditions associated
with the adjoint wavevector $\bm{k}^{\dagger}_{\perp} = -\bm{k}_{\perp}$.
This is convenient for the FEM programming since the same subprograms can be used
to handle the partial differential equations~(\ref{primal:mode:1})
and (\ref{adjoint:mode:2}) while only a few lines of code are needed to
manage the sign change for $\bm{k}_{\perp}$.
%

%
The spectral theory for non-self-adjoint operators is difficult and in general less developed.
In this paper we assume that the modes $\bm{E}_{n}$ form a complete set and
that the adjoint modes $\bm{E}^{\dagger}_n$ can be numbered such that
${\zeta^{\dagger}}_n^2 = \zeta_n^2$, and the following biorthogonality relationship is satisfied

\begin{eqnarray}
\label{eq1:biorthog}
\int_{\Omega} \bm{e}_z \cdot ( \bm{E}^{\dagger}_{m} \times
\bm{H}_{n}) dA
& = & \delta_{mn},
\end{eqnarray}
in which $\bm{H} = [\bm{H}_{\perp},H_z]$
can be obtained from the electric field $\bm{E} = [\bm{E}_{\perp}, E_z]$
using the relation
\begin{eqnarray}
\bm{H} & = & \frac{\nabla \times \bm{E}}{i \, k \, \mu }
\end{eqnarray}
which is derived directly from Maxwell's equations~(\ref{a2}).

A similar spectral property has been proven for a class of non-self-adjoint Sturm-Liouville problems
(see for instance, Theorem 5.3 of Ref.~\cite{Hanson:Springer}).
It is not clear if such a theorem can be extended to the vectorial waveguide mode problem, although
it has been shown that the spectral theory of compact operators
can be applied to a waveguide problem~\cite{Vardapetyan:MC:2003}.
However, our numerical calculations have generated modes which satisfy the biorthogonality relation
and verify the completeness relations Eqs~(\ref{MC5}) and (\ref{ip10}), in Appendix~\ref{MC},
to generally within $10^{-4}$, and with even a better convergence when the number of plane wave orders
and array modes, used in the truncated expansion, increase.

We now establish the biorthogonality relation
{Eq.}~(\ref{eq1:biorthog})
using the operator definitions
{Eqs.}~(\ref{def:adjoint:L}) and (\ref{def:adjoint:M})
of the adjoint, and Eqs (\ref{primal:mode:1}) and (\ref{adjoint:mode:2}).  This
relation may also be established directly from
Maxwell's equations, as is shown in Appendix~\ref{OM}.

We begin with
\begin{eqnarray}
\left( \mathcal{L}^{\dagger} \overline{\bm{E}^{\dagger}_m} , \bm{E}_n \right)
& = &
\left( \overline{\bm{E}^{\dagger}_m} , \mathcal{L} \, \bm{E}_n \right),
\end{eqnarray}
and, since $\bm{E}_n$ and $\bm{E}^{\dagger}_m$ are eigenfunctions, we obtain
\begin{eqnarray}
\left( \overline{\zeta^{\dagger}_m}^2 \mathcal{M}^{\dagger} \overline{\bm{E}^{\dagger}_m} , \bm{E}_n \right)
& = &
\left( \overline{\bm{E}^{\dagger}_m} , \zeta_n^2 \mathcal{M} \, \bm{E}_n \right).
\end{eqnarray}
Now, by taking into account Eq.~(\ref{def:adjoint:M}), we can derive
the following biorthogonality property:
\begin{eqnarray}
\left( \zeta_n^2 - {\zeta^{\dagger}_m}^2 \right)
\left( \overline{\bm{E}^{\dagger}_m} , \mathcal{M} \, \bm{E}_n \right)
& = & 0 \, ,
\end{eqnarray}
i.e.,
\begin{eqnarray}
\label{eq2:biorthog}
\left( \overline{\bm{E}^{\dagger}_m} , \mathcal{M} \, \bm{E}_n \right)
& = & 0, \quad \mbox{if $\zeta_n^2 \neq {\zeta^{\dagger}_m}^2$} \, .
\end{eqnarray}
The integrand of the field product in Eq.~(\ref{eq2:biorthog}) can be expressed
in term of the fields $\bm{H}_{m}$ and $\bm{E}^{\dagger}_{n}$ as in Eq.~(\ref{eq1:biorthog})
by noting
\begin{eqnarray}
 \bm{e}_z \cdot ( \bm{E}^{\dagger}_{m} \times \bm{H}_{n})
& = &
\frac{
\bm{E}^{\dagger}_{m, \perp} \cdot
\left( i \, \zeta_n \, \bm{E}_{n,\perp}
-
\nabla_{\perp} E_{n,z}
\right) }
{i \, k \, \mu } \\
& = &
- \frac{
\zeta \,
\bm{E}^{\dagger}_{m,\perp} \cdot
\left( \mathcal{M} \, \bm{E}_n \right)_{\perp}
}
{k},
 \\
& = &
- \frac{
\zeta \,
\bm{E}^{\dagger}_{m} \cdot
\left( \mathcal{M} \, \bm{E}_n \right)
}
{k},
\end{eqnarray}
since $\left( \mathcal{M} \, \bm{E}_n \right)_{z} = 0$, (\ref{primal:mode:1}) and (\ref{operator:L}).

The biorthogonality relation
{Eq.}~(\ref{eq2:biorthog})
is useful for the FEM implementation of the field product
since the product of two vectors $\bm{v}$ and $\bm{v}^{\dagger}$ takes the
form $- \zeta \, (\bm{v}^{\dagger} \cdot (\bm{M} \, \bm{v}))/k$ where $\bm{M}$
is the discrete version of the operator $\mathcal{M}$ and
is the matrix on the right hand side of Eq.~(\ref{matrix:sym2}).

The calculation of the modes $\bm{E}_n$ and the adjoint modes $\bm{E}^{\dagger}_n$
 using this FEM approach is highly efficient and numerically stable.

In the following section we use modes of the structure for the field
expansion inside  the photonic crystal slab, and exploit the adjoint modes, which
are biorthogonal to the primal modes, in the solution of the field
matching problem in a least square sense.

%
%

\subsection{Fresnel interface reflection and transmission matrices}
\label{FM}

In this section we introduce the Fresnel reflection and transmission
matrices for
{photonic crystal-air}
interfaces and calculate the total transmission,
reflection and absorption of a  photonic crystal slab. First we
introduce the Fresnel reflection $\bm{R}_{12}$ and transmission
$\bm{T}_{12}$ matrices for an interface between free space and the
semi-infinite array of cylinders. We specify an incident plane wave
field $\bm{f}^{E/M-}$ (see Eqs(\ref{epw})--(\ref{kpw})) propagating
from above onto a semi-infinite slab, giving rise to an upward
reflected plane wave field $\bm{f}^{E/M+}$ and a downward
propagating field of modes $\bm{c}^-$ in the slab.

The field matching equations between the plane wave expansions Eqs~(\ref{epw}) and (\ref{kpw})
and the array mode expansion
{Eq.}~(\ref{array:mode:expansion})
are obtained by enforcing
the continuity of the tangential components of transverse fields on either side of the interface:
\begin{eqnarray}
\sum_{s}
\chi^{-1/2}_s \left(f^{E-}_s  +  f^{E+}_s \right) \bm{R}^E_s  +
\chi^{1/2}_s \left(f^{M-}_s + f^{M+}_s \right)  \bm{R}^M_s \nonumber \\
   = \; \sum_{n}  c_n^-  \bm{E}_{n\perp},
\hspace{2cm}
\label{eq12a} \\
\sum_{s}
\chi^{1/2}_s \left(f^{E-}_s - f^{E+}_s \right) \bm{R}^E_s  +
\chi^{-1/2}_s \left(f^{M-}_s - f^{M+}_s \right)  \bm{R}^M_s \nonumber\\
= \; \sum_{n}  c_n^- \left(\bm{e}_z \times \bm{H}_{n\perp} \right).
\hspace{2cm}
\label{eq12b}
\end{eqnarray}
Equations~(\ref{eq12a}) and (\ref{eq12b}) correspond to the continuity condition of the tangential electric
field and  magnetic field respectively. Here $\bm{E}_{n\perp}$
denotes the
{downward}
tangential electric field component of mode $n$ while
$\bm{H}_{n\perp}$ denotes the
{downward}
tangential magnetic field of mode $n$,
{
which satisfy the orthonormality relation
\begin{eqnarray}
\int_{\Omega} \bm{e}_z \cdot ( \bm{E}^{\dagger}_{m} \times
\bm{H}_{n}) dA
& = & - \delta_{mn}.
\end{eqnarray}
}
In Appendix \ref{MC} we derive completeness relations for both the Bloch modes basis
and the plane wave basis.

We now proceed to solve these equations in a least squares sense, using the
Galerkin-Rayleigh-Ritz method in which the two sets of equations are
respectively projected on the two sets of basis functions. In this treatment,
we project the electric field equation onto the plane wave basis and the
magnetic field equation onto the slab  mode basis to derive, in matrix form,
\begin{eqnarray}
\bm{X}^{-1/2} \left(\bm{f}^- + \bm{f}^+ \right) &=& \bm{J}  \bm{c}^-,\label{eqe12}\\
{\bm J}^\dag  \bm{X}^{1/2} \left(\bm{f}^- - \bm{f}^+ \right) &=&  \bm{c}^-,
\label{eqk12}
\end{eqnarray}
where
\begin{eqnarray}
\bm{f}^\pm & = &  \left(\begin{matrix} \bm{f}^{E\pm} \\ \bm{f}^{M\pm} \end{matrix}
\right), \,\,\, \bm{X} = \left(\begin{matrix} \bm{\chi} & \bm 0 \\ \bm{0} &
\bm{\chi}^{-1} \end{matrix} \right), \,\,\,  \\
\bm{J} & = &  \left( \begin{matrix} \bm{J}^E \\ \bm{J}^M \end{matrix} \right),
\qquad \bm{J}^{E/M} = \left[ J^{E/M}_{sm}  \right],\nonumber\\
 \qquad J^{E/M}_{sm} & = & \iint
\overline{\bm{R}}^{E/M}_s \cdot \bm{E}_{m\perp} \,\, dS, \\
\bm{J}^\dag & = &  \left( \begin{matrix} {\bm{J}}^{\dag E} \,\, \bm{J}^{\dag M} \end{matrix} \right),
\qquad \bm{J}^{\dag E/M} = \left[ J^{\dag E/M}_{ms}  \right],\nonumber\\
 \qquad J^{\dag E/M}_{ms} & = & \iint
{\bm{R}}^{E/M}_s \cdot \bm{E}^{\dag}_{m\perp} \,\, dS,
\label{JMAT} \\
\,\,\bm{\chi} & = &
\mathrm{diag~} \{ \chi_s \}.
\end{eqnarray}
Then, defining the scattering matrices $\bm{R}_{12}$ and $\bm{T}_{12}$ according to
the definitions
\begin{equation}
\bm{f}^+ = \bm{R}_{12} \bm{f}^-, \qquad \bm{c}^- = \bm{T}_{12} \bm{f}^-,
\end{equation}
we may solve Eqs (\ref{eqe12}) and (\ref{eqk12}) to derive
\begin{eqnarray}
\bm{R}_{12} = -\bm{I} + 2 \bm{A}  \left( \bm{I} + \bm{B A} \right)^{-1} \bm{B}
= (\bm{AB} + \bm{I})^{-1} (\bm{A}\bm{B} - \bm{I}), \label{eqR12}
\end{eqnarray}
\begin{eqnarray}
\bm{T}_{12} &=& 2 \left( \bm{I} + \bm{BA} \right)^{-1}  \bm{B},
\label{eqT12}  \\
{\rm where~~} \bm{A} & =& \bm{X}^{1/2} \bm{J}, \,\,\bm{B}=\bm{J}^\dag \bm{X}^{1/2}.
\end{eqnarray}

For a structure with inversion symmetric inclusions the following
useful relations hold: $\bm{E}^\dag_n(\bm{r})=\bm{E}_n(-\bm{r})$,
$\bm{J}^\dag=\bm{J}^T$ and $\bm{B}=\bm{A}^T$. While the Fresnel
matrices $\bm{R}_{12}$ and $\bm{T}_{12}$ given in Eqs (\ref{eqR12})
and (\ref{eqT12}) have been derived by presuming a plane wave field
incident from above, identical forms are derived if we presume
incidence from below, a consequence of the given symmetries of the
modes.

We now derive the slab-free space Fresnel coefficients, assuming
that we have a modal field $\bm{c}^-$ incident from above and giving
rise to a reflected modal field $\bm{c}^+$ and a transmitted plane
wave field $\bm{f}^-$ below.  This time, the field matching
equations are
\begin{eqnarray}
\sum_{s} \chi^{-1/2}_s f^{E-}_s  \bm{R}^E_s + \chi^{1/2}_s
f^{M-}_s   \bm{R}^M_s &=& \nonumber\\
\sum_{n}  \left(c_n^- + c_n^+
\right) \bm{E}_{n\perp},
\\
\sum_{s} \chi^{1/2}_s f^{E-}_s \bm{R}^E_s
+
\chi^{-1/2}_s f^{M-}_s \bm{R}^M_s &=&\nonumber\\
\sum_{n}  \left(c_n^- - c_n^+ \right)
\left(\bm{e}_z \times \bm{H_}{n\perp} \right),
\end{eqnarray}
and we again project the electric field equation onto the plane wave basis and
the magnetic field equation on to the modal basis for the slab.  Accordingly,
we derive
\begin{eqnarray}
\bm{X}^{-1/2} \bm{f}^- &=& \bm{J}  \left( \bm{c}^- +  \bm{c}^+ \right), \label{eqe21}\\
 \bm{J^\dag} \bm{X}^{1/2} \bm{f}^- &=&    \bm{c}^- -\bm{c}^+.\label{eqk21}
\end{eqnarray}
Then, defining the scattering matrices $\bm{R}_{21}$ and $\bm{T}_{21}$ according
to $\bm{c}^+ = \bm{R}_{21} \bm{c}^-$ and $\bm{f}^+ = \bm{T}_{21} \bm{c}^-$, we
form
\begin{eqnarray}
\bm{R}_{21} &=& \left( \bm{I} - \bm{BA} \right) \left( \bm{I} + \bm{BA}\right)^{-1}, \label{eqR21}\\
\bm{T}_{21} &=& 2 \bm{A}  \left( \bm{I} + \bm{BA} \right)^{-1}.
 \label{eqT21}
\end{eqnarray}
These Fresnel scattering matrices $\bm{R}_{12}$, $\bm{T}_{12}$,
$\bm{R}_{21}$, $\bm{T}_{21}$ can now be readily utilized to
calculate the total transmission, reflection and the absorption of
the slab. Note that for inversion symmetric inclusions the
following reciprocity relations hold: $\bm{T}_{12}^T=\bm{T}_{21}$
and $\bm{R}_{12}^T=\bm{R}_{21}$. These relations hold independently
of the truncation of the field expansions. In addition, in general,
the following relations are also true:
$\bm{T}_{12}\bm{T}_{21}=\bm{I}-\bm{R}_{{21}}^2$
and
$\bm{T}_{21}\bm{T}_{12}=\bm{I}-\bm{R}_{{12}}^2$.

%
%

\subsection{Transmission, reflection of the slab}
\label{slabTRA}

\begin{figure}
\centerline{\includegraphics[width=5.5cm]{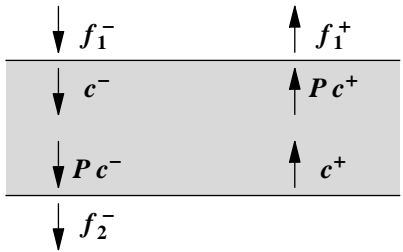}}
\caption{Schematic of the field representation near the top and bottom interfaces.
}
\label{Fig:diagram}
\end{figure}

%
Following the diagram in Fig.~\ref{Fig:diagram}, we may
use the Fresnel interface matrices to calculate the
transmission and the reflection matrices for the entire structure
from the following equations:
\begin{align}
\label{SF1}
\bm{f}^{+}_1 &= \bm{R}_{12} \bm{f}^{-}_1 + \bm{T}_{21} \bm{P} \bm{c}^{+},\\
\label{SF2}
\bm{c}^{-} &= \bm{T}_{12} \bm{f}^{-}_1 + \bm{R}_{21} \bm{P} \bm{c}^+,\\
\label{SF3}
\bm{c}^{+} &= \bm{R}_{21} \bm{P} \bm{c}^{-},\\
\label{SF4}
\bm{f}^{-}_2 &= \bm{T}_{21} \bm{P} \bm{c}^{-},
\end{align}
where $\bm{P} = {\rm diag \, }[\exp(i \, \zeta_n \, h)]$ is the
diagonal matrix which describes the propagation of the $n^{th}$
Bloch mode inside the slab with a thickness $h$. The transmission
and reflection  matrices defined as $\bm{f}_2^-=\bm{T}\bm{f}_1^-$,
$\bm{f}_1^+=\bm{R}\bm{f}_1^-$ can be deduced from the system
{Eqs.}~(\ref{SF1})--(\ref{SF4})
as
\begin{align}
\label{TR}
\bm{T} &= \bm{T}_{21} \bm{P} (\bm{I}-\bm{R}_{21} \bm{P} \bm{R}_{21}\bm{P})^{-1}\bm{T}_{12},\\
\label{TRR}
\bm{R} &= \bm{R}_{12} + \bm{T}_{21} \bm{P} (\bm{I}-\bm{R}_{21} \bm{P} \bm{R}_{21}\bm{P})^{-1} \bm{R}_{21} \bm{P} \bm{T}_{12}.
\end{align}
The amplitudes of the transmitted and reflected fields are then
given by, $\bm{t}=\bm{T}\boldsymbol{\delta}$,
$\bm{r}=\bm{R}\boldsymbol{\delta}$, where
$\bm{\delta}=[0, \ldots, 0, \cos\delta, 0, \ldots, 0, \sin\delta, 0, \ldots,0]^T$ is the
vector containing the magnitudes of components of the incident plane
wave in the specular diffraction order, and $\delta$ is the angle between the electric vector and the plane
of incidence. The absorptance, $A$, is calculated by energy
conservation as
\begin{align}
\label{Absorption}
A = 1 - \displaystyle\sum_{s \in \mathcal{P}} [|r_s|^2 + |t_s|^2],
\end{align}
where $r_s$, $t_s$ are the diffraction order components of $\bm{r}$, $\bm{t}$ and $\mathcal{P}$ is the set of all propagating orders in free space.

In the absence of absorption the following relations for the slab
reflection $\bm{R}$ and transmission $\bm{T}$ matrices can be
deduced (for details see Appendix  C). Given that the photonic
crystal slab is up/down symmetric, the slab transmission $\bm{T}'$
and the reflection $\bm{R}'$ matrices for plane wave incidence from
below are the same as for incidence from above:  $\bm{T}'=\bm{T}$
and $\bm{R}'=\bm{R}$. Therefore the energy conservation relations
take the form
\begin{eqnarray}
\bm{R}^\text{H}\bm{I}_1\bm{R} + \bm{T}^\text{H}
\bm{I}_1\bm{T} & \!\!\! = & \!\!\bm{I}_1
+ i \bm{R}^\text{H}\bm{I}_{\overline{1}}
- i \bm{I}_{\overline{1}}\bm{R} , \label{b37} \\
\bm{R}^\text{H}\bm{I}_1\bm{T} + \bm{T}^\text{H}
\bm{I}_1\bm{R} & \!\!\! = & \!\!i \bm{T}^\text{H}\bm{I}_{\overline{1}}
- i \bm{I}_{\overline{1}}\bm{T} , \label{b39}
\end{eqnarray}
with $\bm{I}_1$ denoting a diagonal matrix with the entries $1$ for the propagating
plane wave channels and zeros for the evanescent plane wave channels, and
$\bm{I}_{\overline{1}}$ being a diagonal matrix which has  entries
{$+ 1$}
in the evanescent $TE$ channels and
{$- 1$}
in  the evanescent $TM$  channels, and zeros for the propagating channels.

The semi-analytic expressions for the transmission
{Eq.}~(\ref{TR})
and the reflection
{Eq.}~(\ref{TRR})
matrices for the slab can give important
insight to improve the overall absorption efficiency.
For instance, in Eq.~(\ref{TR})
the matrices $\bm{T}_{12}$ and $\bm{T}_{21}$ represent the coupling matrices
for a plane wave into and out of the slab, while the scattering matrix
\begin{equation}
(\bm{I}-\bm{R}_{21} \bm{P} \bm{R}_{21}\bm{P})^{-1}
\label{FPM}
\end{equation}
describes Fabry-Perot-like multiple scattering. As demonstrated in
Ref. \cite{Sturmberg:OE:2011}, absorption is enhanced if first,
there is a  strong coupling ($\bm{T}_{12}$), strong scattering
amplitudes
{Eq.}~(\ref{FPM})
that increase the effective path in the slab
multiple times,  and the field strength is concentrated in the
region of high absorption.

%
%

\section{Numerical simulations and verifications}
\label{Applica}

In this section, we first use our mode solver to compute the dispersion curves
of an array of lossless cylinders.
Then we apply our modal approach to analyze the absorption spectrum of an array of lossy cylinders (silicon)
and we also examine the convergence of the method with respect to the truncation parameters.
Finally, we consider the example of a photonic crystal slab which exhibits Fano resonances.

%
%
\subsection{Dispersion curves of an array of cylinders}
\label{Section:Dispersion}
\begin{figure}
\centerline{\includegraphics[width=8.0cm]{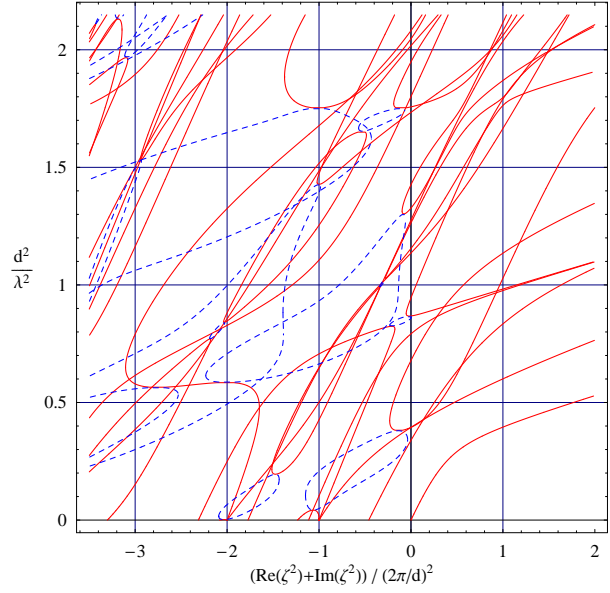}}
\caption{Dispersion curves (lossless cylinders).
The solid red curves represent solutions such that $\zeta^2$ is real.
The dashed blue curves represent solutions such that ${\rm Im}(\zeta^2)$ is not zero.
The complex solutions $\zeta^2$ occur as conjugate pairs and,
in order to differentiate the pairs, we use the term
${\rm Re}(\zeta^2) + {\rm Im}(\zeta^2)$ for the x-axis instead of ${\rm Re}(\zeta^2)$.
}
\label{fig:dispersion:1}
\end{figure}

Though our method can be applied to inclusions of any cross section,
we first consider an array of lossless circular cylinders, with dielectric constant
$n_c^2=8.9$ (alumina), and normalized radius $a/d = 0.2$, in an air background
(refractive index $n_b=1$).
We compute the propagation constant $\zeta$ of the Bloch modes
defined in Eq.~(\ref{mode1}) using the vectorial FEM.
Figure~\ref{fig:dispersion:1} shows the dispersion curves
corresponding to a periodic boundary condition in the transverse
plane, i.e., $\bm{k}_{\perp} = (\alpha_0, \beta_0) =(0,0)$. The solid red
curves indicate values of the propagation constant $\zeta$ such that
$\zeta^2$ is real---with positive values of $\zeta^2$
corresponding to propagating modes, while negative values indicate
evanescent modes.

The dispersion curve for the fundamental propagating mode
 is at the lower right corner and starts at the coordinate origin.
Complex values of $\zeta^2$ can also occur, even
for a lossless system, and even with ${\rm Re} \, \zeta^2 >0 $;
these modes, which occur in conjugate pairs, are
shown by the dashed blue curves and are
distinguished by a horizontal axis which is labeled  ${\rm Re} \, \zeta^2 + {\rm Im} \, \zeta^2$.
We can observe that
the dashed blue curves connect a maximum point of a solid red curve to a minimum
point of another solid red curve. This property of the dispersion of
cylinders arrays was observed by Blad and Sudb{\o}~\cite{Blad:OE:2009}.

The dispersion curves in
Fig.~\ref{fig:dispersion:1} are plotted using the same parameters as
in Fig.~4 of Ref.~\cite{Blad:OE:2009}. All curves shown in Fig.~4 of
\cite{Blad:OE:2009} also appear in Fig.~\ref{fig:dispersion:1} of
the present paper although our figure reveals many additional
curves, for instance, near ($({\rm Re} \, \zeta^2 + {\rm Im} \,
\zeta^2)/(2 \, \pi/d)^2, d^2/\lambda^2) = (-2., 0.1)$. In
\cite{Blad:OE:2009}, the dispersion curves were plotted using a
continuation method whose starting points are on the axis $\zeta=0$;
most of the curves which do not intersect this axis are missing in
Fig.~4 of \cite{Blad:OE:2009}
{but their solutions are required for the completeness of the}
modal expansion Eqs~(\ref{eq12a}) and (\ref{eq12b}).
For instance, if the eigenvalues $\zeta_n$ are numbered in
decreasing order of ${\rm Re} \, \zeta_n^2$, then the index number of the
eigenvalues, which appear near ($({\rm Re} \, \zeta^2 + {\rm Im} \,
\zeta^2)/(2 \, \pi/d)^2, d^2/\lambda^2) = (-2.0, 0.1)$ in
Fig.~\ref{fig:dispersion:1}, are between 10 and 20 and, as explained
in the convergence study of the next section, we typically need well
over 20 modes to obtain good convergence.
%

%
%
\subsection{Absorptance of a dilute silicon nanowire array}
\begin{figure}
\centerline{\includegraphics[width=8.5cm]{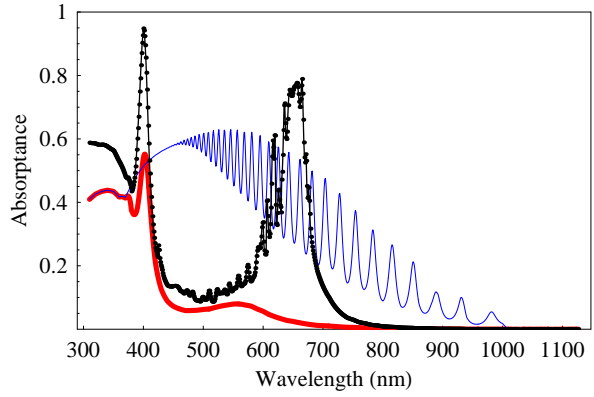}}
\caption{Dotted black curve is the absorption spectrum at normal
incidence for a dilute SiNW array (silicon fill fraction is approximately 3.1\%).
For comparison, the absorptance curves of a homogeneous slab of equal thickness
(thin blue curve) and of a homogeneous slab comprising equal volume
of silicon (thick red curve) are shown.}
\label{fig:dilute:SiNW:1}
\end{figure}

We now consider a silicon nanowire (SiNW) array consisting of absorptive
nanowires of radius $a = 60 \, {\rm nm}$, arranged in square lattice of
lattice constant $d = 600  \, {\rm nm}$.
This constitutes a dilute SiNW array since the silicon fill fraction is approximately 3.1\%.
The dilute nature of the array can facilitate the identification of the modes which
play a key role in the absorption mechanism~\cite{Sturmberg:OE:2011}.
The height of the nanowires is $h = 2.33 \, \mu{\rm m}$.
For silicon we use the complex refractive index of Green and
Keevers~\cite{Green:PPRA:1995}.
Figure~\ref{fig:dilute:SiNW:1} shows the absorptance spectrum of the dilute SiNW array,
together with the absorptance
of a homogeneous slab of equivalent thickness and
of a homogeneous slab of equivalent volume of silicon.
The absorption feature between 600 and 700~nm is absent in bulk
silicon and is entirely due to the nanowire geometry. Using our
method we have identified some specific Bloch modes which play a key
role in this absorption behavior~\cite{Sturmberg:OE:2011}.
At shorter wavelengths the absorption of the silicon is high and therefore
the absorption of the slab does not depend on the slab thickness
(see Fig.\ref{fig:dilute:SiNW:1} thin blue curve and thick red curve).

\begin{figure}
\centerline{\includegraphics[width=8.5cm]{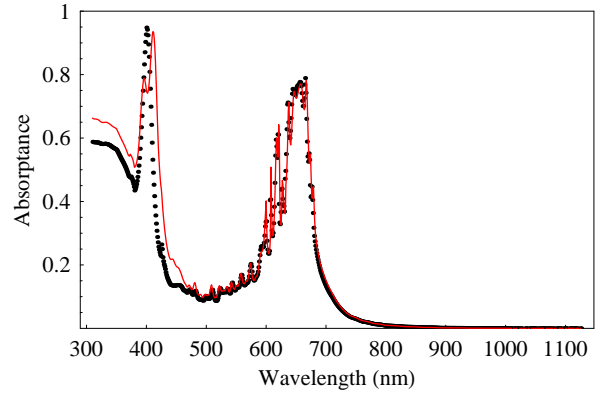}}
\caption{The black dotted curve and the red solid curve are
the absorption spectrum, at normal incidence, of dilute SiNW arrays
consisting respectively of circular cylinders with radius $a = 60 \, {\rm nm}$
and square cylinders with side length $a \, \sqrt{\pi}$.
The two types of cylinders have the same cross section area and the same height.
}
\label{fig:dilute:SiNW:2}
\end{figure}

Note that that the geometry of the inclusion does not need to be circular
since our FEM based method can handle arbitrary inclusion shapes.
Indeed, in Fig.~\ref{fig:dilute:SiNW:2} the absorptance spectrum of a SiNW array
consisting of square cylinders is analyzed and compared to the absorptance for
circular cylinders of
{same period and cross sectional area.}
%
At long wavelengths, the absorption for the two geometries is the same,
while at  shorter wavelengths the absorption is
{slightly}
higher for the square cylinders.
This can be explained by the field concentration
at the corners of the square cylinders~\cite{Meixner}.

\begin{figure}
\centerline{\includegraphics[width=8.5cm]{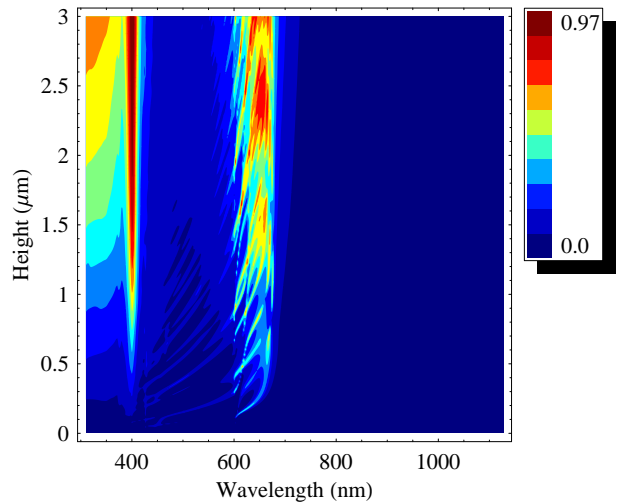}}
\caption{Dilute SiNW array: Contour plot of the absorptance as a function of the wavelength and
the cylinder height $h$.
The cylinder radius and the lattice constant are respectively $a = 60 \, {\rm nm}$ and
$d = 600  \, {\rm nm}$.
}
\label{fig:dilute:SiNW:3}
\end{figure}

The contour plot in Fig.~\ref{fig:dilute:SiNW:3} shows the
absorptance versus wavelength and the cylinder height $h$ for the
%
{circular}
SiNW array.
Note that the nanowire height of $h = 2.33 \, \mu{\rm m}$ used in
Fig.~\ref{fig:dilute:SiNW:1} is, indeed, in a region of high
absorptance for the wavelength band $[600 \, {\rm nm}, 700 \, {\rm nm}]$.
Note that the propagation matrix $\bm{P}$ is the only matrix in the
expressions~(\ref{TR}) and (\ref{TRR})
for the reflection and transmission matrices which depends on the
thickness $h$ and it can be easily updated when $h$ is varied for a
fixed value of the wavelength. Thus, the modal method is a very fast
technique in exploring the dependence of the absorptance with respect to
the thickness. Indeed, the contour plot is obtained by computing the
absorptance for 6001 height values uniformly spaced over the
interval $[0 \, \mu{\rm m}, 3 \, \mu{\rm m}]$, for 409
wavelength values uniformly distributed over the interval
$[310 \, {\rm nm}, 1126 \, {\rm nm}]$; a high sampling resolution is required
in order to capture the oscillatory features which occur in the
region $\lambda \in [400 \, {\rm nm}, 700 \, {\rm nm}]$.
It took about 44 hours to generate the full results
using 16 cores of a high performance parallel computer with 256 cores
(it is a shared memory system consisting of 128 processors Intel Itanium 2 1.6GHz (Dual Core)).
If the absorptance had to be computed independently for the $6001
\times 409$ data points, this would have required many months of
computer time.

We have studied the convergence with respect to the truncation
parameters of the plane wave expansions and array mode expansions in
Eqs~(\ref{eq12a}) and (\ref{eq12b}).
The array modes are ordered in decreasing order with respect to ${\rm Re } \, \zeta_n^2$.
We have used a {\em circular truncation} for the plane wave truncation
number $N_{\rm PM}$, i.e., for a given value of $N_{\rm PM}$, only
the plane wave orders $(p,q)$ such that $p^2+q^2 \leq N_{\rm PM}^2$
are used in the truncated expansions; this choice is motivated by
the fact that, for normal incidence, it is consistent with the
ordering of the array modes since the plane wave propagation
constants are given by the dispersion relation
$\gamma_s^2 = k^2 - (p^2+q^2) \, (2 \pi/d)^2$
(see Eq.~(\ref{eq1:PW:dispersion})); the propagation constants $\gamma_s$
are also numbered in decreasing value of $- (p^2+q^2)$.

\begin{figure}
\centerline{\includegraphics[width=8.5cm]{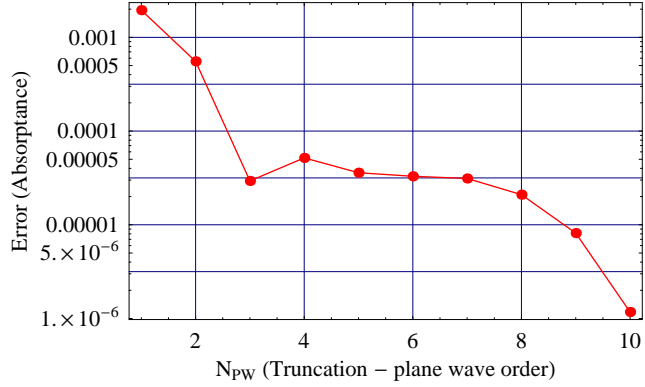}}
\caption{Dilute SiNW array: Convergence as the plane wave truncation number $N_{\rm PM}$ increases for a
fixed wavelength $\lambda = 700 \, {\rm nm}$ and waveguide truncation number $N_{\rm array} = 150$.
The computed absorptance for $N_{\rm PM} = 10$ and $N_{\rm array} = 160$ is $A=0.13940$
and this value is considered as ``exact'' and used to compute the error curve.}.
\label{fig:dilute:SiNW:4}
\end{figure}

\begin{figure}
\centerline{\includegraphics[width=8.5cm]{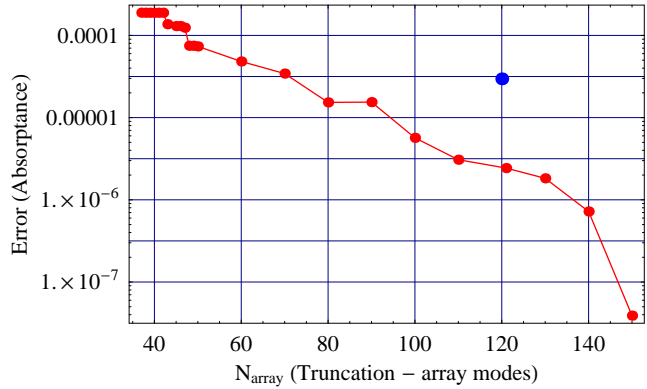}}
\caption{Dilute SiNW array:
Convergence as the array mode truncation number $N_{\rm array}$ increases.
The details given in the caption of Fig.~\ref{fig:dilute:SiNW:4} also apply here
except that the plane wave truncation number is fixed at $N_{\rm PM} = 10$.
Note that $\zeta_{120} = \zeta_{121}$ is a degenerate eigenvalue.
The isolated blue dot corresponds to the truncation value  $N_{\rm array} = 120$
where the error is unexpectedly high;
thus the truncation $N_{\rm array} = 121$ is instead used for the error curve.
}
\label{fig:dilute:SiNW:5}
\end{figure}

Figures~\ref{fig:dilute:SiNW:4} and \ref{fig:dilute:SiNW:5}
illustrate the convergence  when the number of plane wave orders and the
number of array modes used in modal expansions are increased. The
wavelength is set to $\lambda = 700 \, {\rm nm}$ and the
corresponding silicon refractive index is $n = 3.774 + 0.011 \, i$,
which is taken from Ref.~\cite{Green:PPRA:1995}. The error is
estimated by assuming that the result obtained with the highest
discretization is ``exact" ($A=0.13940$ for $N_{\rm PM} = 10$ and
$N_{\rm array} = 160$).
The calculations are based on a highly refined FEM mesh consisting
of 8088 triangles and 16361 nodes.
In Fig.~\ref{fig:dilute:SiNW:5}, there is a
sudden jump in error when $N_{\rm array} = 120$ (isolated blue dot);
this is due to the
chosen truncation cutting through a pair of degenerate
eigenvalue $\zeta_{120} = \zeta_{121}$ i.e., including one member of the pair but excluding the other).
Indeed the solution is well-behaved when $N_{\rm array} = 121$.
Similar behavior has been
observed when a pair of conjugate eigenvalues is cut. Thus all
members of a family of eigenvalues must be included together in
the modal expansion, otherwise there is degradation of the convergence,
which is particularly strong when it occurs for a low-order
eigenvalue which  makes a significant contribution to the modal
expansions. In practice we expect the computed absorptance to have
about three digits of accuracy when the truncation parameters
of the plane wave expansions and array mode expansions are set
respectively to $N_{\rm PM} = 3$ (giving 29 plane wave orders and $2
\times 29$ basis functions for TE and TM polarizations) and $N_{\rm array} = 50$,
assuming adequate resolution of the FEM mesh.
The absorptance curve for the dilute SiNW array in
Fig.~\ref{fig:dilute:SiNW:1} is obtained using these truncation
parameters and an FEM mesh which has 1982 triangles and 4061 nodes.

\begin{figure}
\centerline{\includegraphics[width=8.5cm]{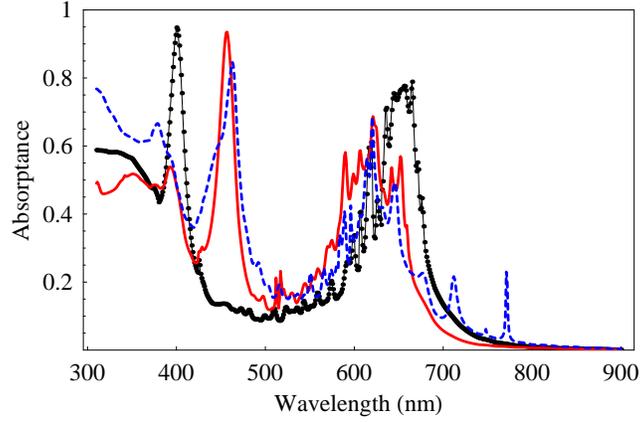}}
\caption{Off-normal incidence on a dilute SiNW: Absorption spectrum
for $45^{\circ}$ off-normal orientated along $x$-axis (azimuthal angle
= 0). The solid red and dashed blue curves represent an incidence by
TE-polarized and TM-polarized plane wave respectively. The
absorption spectrum for normal incidence
(Fig.~\ref{fig:dilute:SiNW:1}) is also shown (dotted black). }
\label{fig:dilute:SiNW:6}
\end{figure}

Figure~\ref{fig:dilute:SiNW:6} presents the absorptance spectrum for
off-normal incidence ($45^{\circ}$). The absorptance is sensitive
to the angle of incidence and the light polarization. Compared with
normal incidence, the absorptance peak in the wavelength band
$[600\, {\rm nm}, 700 \, {\rm nm}]$ (low silicon absorption) has
shifted to shorter wavelengths while the peak near $400 \, {\rm nm}$
(high silicon absorption) has shifted to longer wavelengths.
%

%
%

\subsection{Fano resonances in a photonic crystal slab}

\begin{figure}
\centerline{\includegraphics[width=8.5cm]{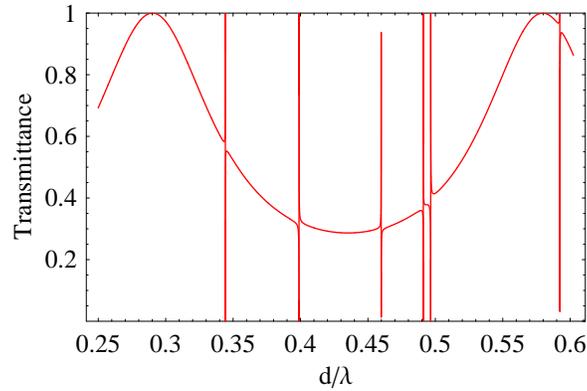}}
\caption{Fano resonances: Transmittance versus normalized frequency for the
normal incidence. The radius of the cylinders is $a/d=0.05$.
The resonances in transmission are well resolved.
}
\label{slabt}
\end{figure}

Fano resonances are well known from the field of particle
physics~\cite{Fano}, and they are observable also in photonic
crystals~\cite{Fan}.
They are notable for their sharp spectral features and so serve as a good benchmark
for the accuracy of new numerical methods.
We have carried out a calculation of Fano resonances using our modal formulation.
We present here an example that was first studied by Fan and Joannopoulos~\cite{Fan}.
The  photonic crystal slab consists of a square array of air holes in a background
material of relative permittivity $\epsilon = 12$.
Figure~\ref{slabt} shows the transmittance of a photonic crystal slab as a function of
the normalized frequency $d/\lambda$ for a plane wave at normal incidence.
The parameters of the slab considered in Fig.~\ref{slabt} are identical to
those in Fig.~12(a) of Ref.~\cite{Fan}, and the curves from the two figures are
the same to visual accuracy.
This is an additional validation of the approach presented here.
The transmittance curve reveals a strong transmission resonance which  is typical of
asymmetric Fano resonances.
These resonances are very sharp and can be used for switching purposes~\cite{Asadi:OC:2011}.

%
%

\section{Conclusion}
\label{Concl}

We have developed a rigorous modal formulation for the diffraction of plane waves by absorbing photonic crystal slabs.
This approach combines the strongest aspects of two methods:
the computation of Bloch modes is handled numerically,
using finite elements in a two-dimensional context where this method excels,
while the reflection and transmission of the fields through the slab interfaces are handled semi-analytically,
using a generalization of the theory of thin films.
This approach can lead to results achieved using a fraction of the computational resources
of conventional algorithms:
an example of this is given in Fig.~\ref{fig:dilute:SiNW:3}, in which a large
number of different computations for different slab thicknesses were able to be computed
in a very short time. Although the speed-testing of this method against conventional methods
(FDTD and 3D Finite Element packages such as COMSOL) is a subject for future
work, we have demonstrated here the method's accuracy and its rapidity of convergence. The method
also satisfies all internal checks related to reciprocity and conservation of energy, as well as reproducing
known results from the literature in challenging situations, such as the simulation of Fano resonances (Fig.~\ref{slabt}).

The method is very general with respect to the geometry of the
structure. We have demonstrated this by modeling both square and circular shaped inclusions.
 In addition,
because the method is at a fixed frequency, it can handle both
dissipative and dispersive structures in a straightforward manner,
using tabulated values of the real and imaginary parts of the
refractive index. This is in contrast to time-domain methods such as
FDTD, or to some formulations of the finite element approach. Though
we have used a single array type here (the square array), other
types of structure (such as hexagonal arrays) can be dealt with by
appropriately adjusting the unit cell $\Omega$, together with the
allowed range of Bloch modes.

It is also easy to see how this method could be extended to multiple
slabs containing different geometries, as well as taking into
account the effect of one or more substrates; this extension would
involve the inclusion of field expansions for each layer, together
with appropriate Fresnel matrices, in the equation system
(\ref{SF1})-(\ref{SF4}). In principle this approach could then be
used to study rods (or holes) whose radius or refractive index
changed continuously with depth, provided the spacing between the
array cells remained constant.

Our method has an important advantage over purely numerical
algorithms in that it gives physical insight into the mechanisms of
transmission and absorption in slabs of lossy periodic media. The
explanation of the absorption spectrum in arrays of silicon nanorods
is vital for the enhancement of efficiency of solar cells, however
this spectrum is complicated, with a number of processes, including
coupling of light into the structure, Fabry-Perot effects, and the
overlap of the light with the absorbing material, all playing an
important role. By expanding in the natural eigenmodes of each layer
of the structure, it is possible to isolate these different effects
and to identify criteria that the structure must satisfy in order to
efficiently absorb light over a specific wavelength range. This, we
have discussed in a recent related
publication~\cite{Sturmberg:OE:2011}.

%
%

\section*{Acknowledgments}

This research was conducted by the Australian Research Council Centre
of Excellence for Ultrahigh Bandwidth Devices for Optical Systems
(project number~CE110001018).
We gratefully acknowledge the generous allocations of computing time
from the National Computational Infrastructure (NCI) and from Intersect Australia.

%
%

\newpage
\appendix

%
%

\section{Modal biorthogonality and normalization}
\label{OM}

Here we prove the biorthogonality of the modes and adjoint modes. Let us consider
set of  modes $(\bm{E}_n,\bm{H}_n)$ and adjoint modes
$(\bm{E}_m^\dag,\bm{H}_m^\dag)$ of photonic crystal. These modes satisfy
\begin{subequations}
\begin{eqnarray}
\nabla \times \bm{H}_n  = & - & i k \varepsilon \bm{E}_n,
\label{ME11}\\
\nabla \times \bm{E}_n  = & & i k \mu \bm{H}_n \label{ME12}
\end{eqnarray}
\end{subequations}
and
\begin{subequations}
\begin{eqnarray}
\nabla \times \bm{H}_m^\dag  = &  & i k \varepsilon \bm{E}_m^\dag,
\label{ME21}\\
\nabla \times \bm{E}_m^\dag  = & -& i k \mu \bm{H}_m^\dag \label{ME22}.
\end{eqnarray}
\end{subequations}
We  multiply both sides of (\ref{ME11}) by $\bm{E}_m^\dag$
and (\ref{ME21}) by $\bm{E}_n$ and correspondingly we multiply each set of  (\ref{ME12}) by $\bm{H}_m^\dag$ and (\ref{ME22}) by $\bm{H}_n$ then adding these
we deduce
\begin{equation}
\nabla \cdot (
\bm{E}_n \times \bm{H}_m^\dag +
\bm{E}_m^\dag \times \bm{H}_n ) = 0.
\label{mainE}
\end{equation}

Next we separate the transverse $\perp$ and longitudinal
$\parallel$ (along the cylinder axes) components according to
\begin{equation}
\bm{E}_n= \bm{E}_{n\perp} +\bm{E}_{n\parallel} , \quad
\bm{H}_n= \bm{H}_{n\perp} +\bm{H}_{n\parallel} , \quad
\nabla= \nabla_\perp + \bm{e}_z\frac{\partial
}{\partial  z}\, . \label{nab}
\end{equation}
After the substitution of (\ref{nab})
into (\ref{mainE}) we obtain
{
\begin{eqnarray}
 \bm{e}_z  \cdot  \frac{\partial}{\partial z}
\left [ \bm{E}_{n\perp}\times \bm{H}^\dag_{m \perp} \right.
 +
\left.
\bm{E}^\dag_{m \perp} \times \bm{H}_{n \perp}\right ] \nonumber \\
 =  \nabla_\perp \cdot \left[ \bm{E}_{n \perp}
\times \bm{H}^\dag_{m \parallel} \right.
 +
\left. \bm{E}_{n \parallel}
\times \bm{H}^\dag_{m \perp} \right.
\nonumber \\
+ \left. \bm{E}^\dag_{m \perp} \times \bm{H}_{n \parallel} \right.
 +
\left.
\bm{E}^\dag_{m \parallel} \times \bm{H}_{n\perp} \right] .
\label{EFO}
\end{eqnarray}
}
Taking into account the $z$-dependence on the modes given by the factors
 $\exp{(i \zeta_n z)}$, $\exp{(-i \zeta_m^\dag z)}$ and integrating
(\ref{EFO}) over the unit cell
we derive
\begin{equation}
i(\zeta_n-\zeta_m^\dag)\int_{\Omega} \bm{e}_z \cdot ( \bm{E}_{n\perp}\times
\bm{H}^\dag_{m \perp} + \bm{E}^\dag_{m \perp} \times
\bm{H}_{n\perp})dA=0, \label{EFOO}
\end{equation}
since the integral on the left hand side vanishes due to quasi-periodicity.
We finally obtain
\begin{equation}
\int_{\Omega} \bm{e}_z \cdot ( \bm{E}_{n\perp}
\times \bm{H}^\dag_{m \perp} + \bm{E}^\dag_{m \perp}
\times \bm{H}_{n \perp})dA=0, \label{EFOOO}
\end{equation}
which holds for arbitrary modes such that $\zeta_n \ne \zeta_m^\dag$.

The same relation holds for the counter propagating mode $m$
\begin{equation}
i(\zeta_n+\zeta_m^\dag)\int_{\Omega} \bm{e}_z \cdot ( \bm{E}^-_{n \perp}
\times \bm{H}^{\dag}_{ n\perp} + \bm{E}^{\dag}_{m \perp} \times
\bm{H}^-_{n \perp})dA=0. \label{EFpm}
\end{equation}
The minus sign in the superscript position indicates the direction of
the propagation.
Taking into account
the relations $\bm{E}^-_{n\perp}=\bm{E}_{n\perp}$ and
$\bm{H}^-_{n\perp}=-\bm{H}_{n\perp}$ we can rewrite (\ref{EFpm}) in
the form
\begin{equation}
\int_{\Omega} \bm{e}_z \cdot ( \bm{E}_{n \perp}
\times \bm{H}^{\dag}_{m \perp} - \bm{E}^{\dag}_{m \perp}\times
\bm{H}_{n\perp})dA=0. \label{EFpmm}
\end{equation}
After  subtraction of relation (\ref{EFpmm}) from (\ref{EFpm})  the
orthogonality relation takes form
\begin{equation}
\int_{\Omega} \bm{e}_z \cdot ( \bm{E}^{\dag}_{m \perp}
\times \bm{H}_{n \perp})dA=0,
\label{orto}
\end{equation}
which states that two distinct modes propagating in the same
direction are orthogonal.
It is then clear that these modes can always be normalized such that
 \begin{equation}
\int_{\Omega}  (\bm{e}_z \times
\bm{H}_{n \perp}) \cdot  \bm{E}^{\dag}_{m \perp}dA=\delta_{mn}.
\label{orto4}
\end{equation}

%
%

\section{Modal Completeness}
\label{MC}

From the field expansions we can derive the condition of the modal completeness.
The plane waves  can be expanded in the following forms:
\begin{eqnarray}
\overline{\bm{R}}^{E/M}_s & = & \sum_n c_n^{E/M}(\bm{e}_z \times \bm{H}^{\dag}_{n \perp}),
\label{MC1} \\
\bm{R}^{E/M}_s & = & \sum_n d_n^{E/M}  \bm{E}_{n \perp}.
\label{MC2}
\end{eqnarray}
By projecting
{Eq.}~(\ref{MC1})
on the modes $\bm{E}_{m \perp}$ and using
the biorthogonality relations
{Eq.}~(\ref{orto4})
we deduce (see Eq.
(\ref{JMAT}))
\begin{eqnarray}
  c_m^{E/M} & = &
\int_{\Omega} \bm{E}_{m \perp} \cdot \overline{\bm{R}}_s^{E/M}\, dA
 = J_{s m}^{E/M} \,.
\label{MC3}
\end{eqnarray}
Similarly we project  (\ref{MC2}) onto the adjoint magnetic mode $(\bm{e}_z \times \bm{H}^{\dag}_{m \perp})$ and deduce
\begin{eqnarray}
  d_m^{E/M} & = &
\int_{\Omega} (\bm{e}_z\times\bm{H}^{\dag}_{m \perp}) \cdot \bm{R}_s^{E/M}\, dA
 = K_{ms}^{E/M} \,.
\label{MC3a}
\end{eqnarray}
Thus,
\begin{eqnarray}
\overline{\bm{R}}^{E/M}_s & = & \sum_n K_{n s}^{E/M}(\bm{e}_z \times \bm{H}^{\dag}_{n \perp}),
\label{MC1A} \\
\bm{R}^{E/M}_s & = & \sum_n J_{n s}^{E/M}  \bm{E}_{n \perp}.
\label{MC2AA}
\end{eqnarray}

Next we project (\ref{MC2}) onto plane wave basis by multiplying both sides of (\ref{MC2})
 by $\overline{\bm{R}}_{s'}^{E/M}$ and integrating over the unit cell. We obtain
\begin{eqnarray}
 \int_{\Omega} \bm{R}^{E/M}_s \cdot \overline{\bm{R}}_{s'}^{E/M}\, dA & = &
\sum_{n} K^{E/M}_{ns'}\int_{\Omega} \bm{E}_{n \perp} \cdot \overline{\bm{R}}_{s'}^{E/M}\, dA \nonumber\\ & = & \sum_{n} K^{E/M}_{ns'}J^{E/M}_{{s'}n}=\delta_{s {s'}}.
\label{MC4}
\end{eqnarray}
The equation (\ref{MC4}) represents the completeness relation for the modes. If we introduce the vectors of matrices
\begin{eqnarray}
\bm{J} = \left[ \begin{array}{l} \bm{J}^E \\ \bm{J}^M \end{array} \right]
\text{ and }
\bm{K} = \left[ \begin{array}{l} \bm{K}^E \\ \bm{K}^M \end{array} \right],
\end{eqnarray}
where
\begin{eqnarray}
\bm{J}^{E/M} = \left[ J_{s m}^{E/M} \right]
\text{ and }
\bm{K}^{E/M} = \left[ K_{s n}^{E/M} \right]
\end{eqnarray}
then the completeness relation
{Eq.}~(\ref{MC4})
can be written  in the matrix form
\begin{eqnarray}
\bm{J} \bm{K } & = & \bm{I},
\label{MC5}
\end{eqnarray}
where $\bm{I}$ is the identity matrix.

The completeness relation of the Rayleigh modes can be established in a similar way.
The transverse component of electric and magnetic modal fields can be represented as
a series in terms of Rayleigh modes  in the region above the grid as
\begin{eqnarray}
\bm{E}_{m \perp} & = & \sum_s \left(
J_{s m}^E \bm{R}_s^E + J_{s m}^M \bm{R}_s^M
\right) ,
\label{ip01} \\
\bm{e}_z \times \bm{H}^{\dag}_{n \perp} & = & \sum_s \left(
K_{ns}^E \overline{\bm{R}}_s^E + K_{ns}^M \overline{\bm{R}}_s^M
\right).
\label{ip02}
\end{eqnarray}
By multiplying (\ref{ip01}) on (\ref{ip02}) and integrating we deduce
\begin{eqnarray}
\int_{\Omega} \bm{E}^{\dag}_{m \perp} \cdot (\bm{e}_z \times \bm{H}_{n \perp}\,)
\,dA & = &
\sum_s \left( J_{s m}^E\, K_{s n}^E + J_{s m}^M\, K_{s n}^M \right)
\nonumber \\
& = & \delta_{n m}.
\label{ip09}
\end{eqnarray}
The completeness relation
{Eq.}~(\ref{ip09})
can be written in matrix form
\begin{equation}
\bm{K} \bm{J} = \bm{I}.
\label{ip10}
\end{equation}

%
%

\section{Interface and slab Energy conservation and reciprocity relations}
\label{ECR}

Here we briefly outline the derivation of energy conservation relations for the
situation when there is no absorption. The flux conservation leads to the certain
 relations between the Fresnel interface reflection and transmission matrices.

For some value $z$ we can write an expansion
\begin{equation}
\bm{E}_\perp=\sum_{n}(c_n^- + c_n^+)\bm{E}_{n\perp}
\label{AIE1}
\end{equation}
and similarly
\begin{equation}
\bm{e}_z\times\bm{H}_\perp=\sum_{n}(c_n^- - c_n^+)\bm{e}_z\times\bm{H}_{n\perp}
\label{AIE2}
\end{equation}
The downward flux is defined by
\begin{eqnarray}
S_z =
\rm{Re} \left[ \int \bm{E}_\perp\cdot(\bm{e}_z\times \overline{\bm{H}}_\perp )\right]
= \hspace{2cm}
\label{AIE3} \\
\frac{1}{2}\left [ (\bm{c}^- - \bm{c}^+)^H\bm{U}(\bm{c}^- +
c^+)+(\bm{c}^-+\bm{c}^+)^H\bm{U}^H(\bm{c}^- - c^+)\right ],
\nonumber
\end{eqnarray}
where the matrix $\bm{U}$ is given by
\begin{equation}
U_{mn}=\int_{\Omega} \bm{E}_{m \perp}\cdot  (\bm{e}_z \times
\overline{\bm{H}}_{n \perp})  dA.
\label{AIE4}
\end{equation}
The relation (\ref{AIE3}) can be written in the form

\begin{eqnarray}
S_z=\rm{Re}\left\{[\bm{c}^{-H}\,\bm{c}^{+H}]\bm{V}\left[
\begin{array}{c}
\bm{c}^-  \\
\bm{c}^+  \\
\end{array}
\right]\right\},
\label{AIE5}
\end{eqnarray}
where matrix
\begin{eqnarray}
\bm{V}=\left[
\begin{array}{ccccccc}
\frac{1}{2}(\bm{U} + \bm{U}^H) &  & & \frac{1}{2}(\bm{U} - \bm{U}^H) &  \\
-\frac{1}{2}(\bm{U} \textcolor{red}{\mbox{\Large\ensuremath $-$}} \bm{U}^H) &  & &
-\frac{1}{2}(\bm{U} \textcolor{red}{\mbox{\Large\ensuremath $+$}} \bm{U}^H) & \\
\end{array}
\right]
\label{AIE6}
\end{eqnarray}
is Hermitian, i.e.  $\bm{V}^H=\bm{V}$.
This is a general result which is applicable also in the presence of
absorption. The $\bm{U}$ matrix is a dense matrix in the presence of
absorption. In the absence of absorption  it reduces to the
following structural form
\begin{eqnarray}
 \bm{U}=\left [
\begin{array}{ccccccccccccccccccccccc}
1 & & 0 & & 0 & & 0    &  & 0    &   & & 0   &    & 0 &    &   &0 &    &   &     & \ldots  & &   0\\
0 & & 1 & & 0 & & 0    &  & 0    &   & & 0   &    & 0 &    &   &0 &    &   &     & \ldots  & &   0\\
0 & & 0 & & 1 & & 0    &  & 0    &   & & 0   &    & 0 &    &   &0 &    &   &     & \ldots  & &   0\\
0 & & 0 & & 0 & & \pm \textcolor{red}{\mbox{\large 1}} &  & 0    &   & & 0   &    & 0 &    &   & 0&    &   &     & \ldots  & &   0\\
0 & & 0 & & 0 & & 0    &  & \pm \textcolor{red}{\mbox{\large 1}} &   & & 0   &    & 0 &    &   & 0&    &   &     & \ldots  & &   0\\
0 & & 0 & & 0 & & 0    &  & 0    &   & & \pm \textcolor{red}{\mbox{\large 1}} &    & 0 &    &   &0 &    &   &     & \ldots  & &   0\\
0 & & 0 & & 0 & & 0    &  & 0    &   & &  0  &    & 0 &    &   &1 &    &   &     &  \ldots        & &   0\\
0 & & 0 & & 0 & & 0    &  & 0    &   & &  0  &    & -1 &    &   &0 &    &   &     &  \ldots        & &   0\\
\vdots & &  & &  & &     &  &     &   & &    &    &  &    &   & &    &   &     &  \ldots        & &   \vdots\\
0 & & \ldots & &  & &     &  &     &   & &    &    &  &    &   & &    &   &     &  0       & &   1\\
\vdots & & \ldots & &  & &     &  &     &   & &    &    &  &    &   & &    &   &     &  -1        & &  0\\
\end{array}
\textcolor{red}{\underline{.}}
\label{AIE8}
\right]
\end{eqnarray}
as we show in Appendix \ref{ECRINT}. Here we have ordered first the
propagating modes with real values of $\zeta_n$, then evanescent
modes with pure imaginary propagating constants $\zeta_n=\pm
i|\zeta_n|$ and finally the evanescent modes with complex valued
propagating constants $\zeta_n=\zeta'_n+i\zeta''_n$. Given the form
of the $\bm{U}$ matrix
{Eq.}~(\ref{AIE8})
the expression for the flux
{Eq.}~(\ref{AIE5})
can be expressed as
\begin{eqnarray}
S_z=[\bm{c}^-\,\bm{c}^+]^H\left[
\begin{array}{ccccccc}
\bm{I}_m &  & & i\bm{I}_{\overline{m}} &  \\
-i\bm{I}_{\overline{m}}&  & & -\bm{I}_m & \\
\end{array}
\right] \left[ \begin{array}{c}
\bm{c}^-  \\
\bm{c}^+  \\
\end{array}
\right],
\label{AIE9}
\end{eqnarray}
where $\bm{I}_m=\bm{U}_s$ - a diagonal matrix with unity on the propagating part of
the diagonal of $\bm{U}$, while $\bm{I}_{\overline{m}}=\bm{U}_a$ corresponds to the
 evanescent part of  $\bm{U}$
{[see Eq.~(\ref{AIE9})].}
%

We next develop the energy conservation relations by considering the
integration of fields at the interface between free space and the
semi-infinite photonic crystal. We write
\begin{eqnarray}
\bm{c}^- & = & \bm{T}_{12}\bm{f}^- + \bm{R}_{21}\bm{c}^+,
\label{AIE10aa}\\
\bm{f}^+ & = & \bm{R}_{12}\bm{f}^-  +  \bm{T}_{21}\bm{c}^+,
\label{AIE10}
\end{eqnarray}
while
\begin{eqnarray}
\left[ \begin{array}{c}
\bm{c}^-  \\
\bm{c}^+  \\
\end{array}
\right]=\left[
\begin{array}{ccccccc}
\bm{T}_{12} &  & & \bm{R}_{21} &  \\
0 &  & & \bm{I} & \\
\end{array}
\right] \left[ \begin{array}{c}
\bm{f}^-  \\
\bm{c}^+  \\
\end{array}
\right].
 \label{AIE10a}
\end{eqnarray}
We also can rewrite the relation (\ref{AIE10}) in the matrix form
\begin{eqnarray}
\left[ \begin{array}{c}
\bm{f}^-  \\
\bm{f}^+  \\
\end{array}
\right]=\left[
\begin{array}{ccccccc}
\bm{I} &  & & 0 &  \\
\bm{R}_{12} &  & & \bm{T}_{21} & \\
\end{array}
\right] \left[ \begin{array}{c}
\bm{f}^-  \\
\bm{c}^+  \\
\end{array}
\right]
 \label{AIE10aaa},
\end{eqnarray}
while the energy flux in the free space as
\begin{eqnarray}
S_z=[\bm{f}^-\,\bm{f}^+]^H\left[
\begin{array}{ccccccc}
\bm{I}_1 &  & & i\bm{I}_{\overline{1}} &  \\
-i\bm{I}_{\overline{1}}&  & & -\bm{I}_1 & \\
\end{array}
\right] \left[ \begin{array}{c}
\bm{f}^-  \\
\bm{f}^+  \\
\end{array}
\right]. \label{AIEngF}
\end{eqnarray}
Now we substitute the relation for modal vector coefficients
$\bm{c}^{\pm}$
{Eq.}~(\ref{AIE10a})
into
{Eq.}~(\ref{AIE9})
and the plane wave coefficients
{Eq.}~(\ref{AIE10aaa})
into
{Eq.}~(\ref{AIEngF}) then by equating
the total fluxes in free space
{Eq.}~(\ref{AIEngF})
and in the photonic crystal
{Eq.}~(\ref{AIE10a})
we can write
\begin{eqnarray}
\left[ \begin{array}{cc}
\bm{I} & \bm{R}_{12}^\text{H} \\
\bm{0} & \bm{T}_{21}^\text{H} \end{array} \right] \left[
\begin{array}{cc}
\bm{I}_1 & -i \bm{I}_{\overline{1}} \\
i \bm{I}_{\overline{1}} & - \bm{I}_1 \end{array} \right] \left[
\begin{array}{cc}
\bm{I} &  \bm{0} \\
\bm{R}_{12} & \bm{T}_{21} \end{array} \right]
\phantom{~~~~~~~~~~~~~} \nonumber \\
= \left[ \begin{array}{cc}
\bm{T}_{12}^\text{H} &  \bm{0} \\
\bm{R}_{21}^\text{H} & \bm{I} \end{array} \right] \left[
\begin{array}{cc}
\bm{I}_2 & -i \bm{I}_{\overline{2}} \\
i \bm{I}_{\overline{2}} & - \bm{I}_2 \end{array} \right] \left[
\begin{array}{cc}
\bm{T}_{12} & \bm{R}_{21} \\
\bm{0} & \bm{I} \end{array} \right] . \label{a36}
\end{eqnarray}
where $\bm{I}_2=\bm{I}_m$ and $\bm{I}_{\overline{2}}=\bm{I}_{\overline{m}}$    and  $\bm{I}_1$
as defined for the plane waves in
Section~\ref{slabTRA}\textcolor{red}{\large\underline{..}}
Expanding these and equating the
formed partitions yields four conservation relations

\begin{eqnarray}
\bm{R}_{12}^\text{H}\bm{I}_1\bm{R}_{12} + \bm{T}_{12}^\text{H}
\bm{I}_2\bm{T}_{12} & \!\!\! = & \!\!\bm{I}_1
+ i \bm{R}_{12}^\text{H}\bm{I}_{\overline{1}}
- i \bm{I}_{\overline{1}}\bm{R}_{12} , \label{a37} \\
\bm{R}_{12}^\text{H}\bm{I}_1\bm{T}_{21} + \bm{T}_{12}^\text{H}
\bm{I}_2\bm{R}_{21} & \!\!\! = & \!\!i \bm{T}_{12}^\text{H}
\bm{I}_{\overline{2}}
- i \bm{I}_{\overline{1}}\bm{T}_{21} , \label{a38} \\
\bm{R}_{21}^\text{H}\bm{I}_2\bm{T}_{12} + \bm{T}_{21}^\text{H}
\bm{I}_1\bm{R}_{12} & \!\!\! = & \!\!i \bm{T}_{21}^\text{H}\bm{I}_{\overline{1}}
- i \bm{I}_{\overline{2}}\bm{T}_{12} , \label{a39} \\
\bm{R}_{21}^\text{H}\bm{I}_2\bm{R}_{21} + \bm{T}_{21}^\text{H}
\bm{I}_1\bm{T}_{21} & \!\!\! = & \!\!\bm{I}_2
+ i \bm{R}_{21}^\text{H}\bm{I}_{\overline{2}}
- i \bm{I}_{\overline{2}}\bm{R}_{21} . \label{a310}
\end{eqnarray}
The slab energy conservation relations can be found in the similar way as for the
interface relations as above. Furthermore the expressions of the energy relations
are very similar  to the interface relations. The only difference is now the
transmission $\bm{T}_{12}$ and the reflection $\bm{R}_{12}$ matrices in (\ref{a310})
need to be replaced by the slab reflection and transmission matrices.

\section{The flux matrix $\bm{U}$ }
\label{ECRINT}

The adjoint modes are defined by
\begin{eqnarray}
\label{mode22}
\mathcal{L}^\dag \, \bm{E}^\dag_n
& = &
\zeta^{2\dag}_n \mathcal{M}^\dag \, \bm{E}^\dag_n,
\end{eqnarray}
with anti-quasi-periodicity condition $\bm{E}^\dag(\bm{r}+\bm{R}_p)=\bm{E}^\dag(\bm{r})\exp{(-i\bm{k}_0\cdot\bm{R}_p)}$
When there is no absorption
\begin{eqnarray}
\label{modeadj}
\mathcal{L} \, \bm{E}^\dag_n
& = &
\zeta^{2\dag}_n \mathcal{M} \, \bm{E}^\dag_n
\end{eqnarray}
given $ \mathcal{L}^\dag=\mathcal{L}$ and $\mathcal{M}^\dag=\mathcal{M}$ are then self adjoint operators.
Note that even though the operators $\mathcal{M}$ and $\mathcal{L}$ are self adjoint (when there
is no absorption) the eigenvalue problem
{Eq.}~(\ref{modeadj})
is not Hermitian because the operator
$\mathcal{M}^{-1}\mathcal{L}$ is not self adjoint in general.
Therefore the eigenvalues can be real representing propagating modes, as well as complex
(pure imaginary or complex) representing evanescent modes.
Now from
{Eq.}~(\ref{modeadj})
we deduce
\begin{eqnarray}
\label{modeadj2}
\mathcal{L} \, \overline{\bm{E}^\dag_n}
& = &
\overline{\zeta^{2\dag}_n} \mathcal{M} \, \overline{\bm{E}^\dag_n},
\end{eqnarray}
where overline means complex conjugation.
By comparing
{Eq.}~(\ref{modeadj2})
and the original eigenvalue equation we deduce
\begin{eqnarray}
\label{modeadj3}
 \bm{E}^\dag_{n\zeta^2}
& = &
 \overline{\bm{E}}_{n\overline{\zeta}^{2 }}
\end{eqnarray}
For the real eigenvalues $\zeta$ we choose $\zeta^\dag=\zeta$ while
for complex $\zeta$ we must choose $\zeta^\dag=-\overline{\zeta}$
which will ensure that
 the downward evanescent propagating field is decaying.  Therefore we have
\begin{eqnarray}
\label{modeadj5}
 \bm{E}^\dag_{n\zeta}
& = &
 \overline{\bm{E}}_{n\overline{\zeta}}
\end{eqnarray}
for the propagating field and
\begin{eqnarray}
\label{modeadj6}
 \bm{E}^\dag_{n\zeta}
& = &
 \overline{\bm{E}}_{n,-\overline{\zeta}}
\end{eqnarray}
for the evanescent field. From Maxwell's equations we deduce that
\begin{eqnarray}
\label{modeadj7}
 \bm{H}^\dag_{\bot n\zeta}
& = &
 -\overline{\bm{H}}_{\bot n,-\overline{\zeta}}.
\end{eqnarray}
Thus, the equations (\ref{EFOO}) and (\ref{EFpm}) can be rewritten as
\begin{eqnarray}
(\zeta_m-\overline{\zeta}_n)\int_{\Omega} \bm{e}_z \cdot ( \bm{E}_{m\perp}\times
\overline{\bm{H}}_{n \perp} + \overline{\bm{E}}_{n \perp} \times
\bm{H}_{m\perp})dA & = & 0,\nonumber \\ \label{EFOOA}\\
(\zeta_m+\overline{\zeta}_n)\int_{\Omega} \bm{e}_z \cdot ( \bm{E}_{m\perp}\times
\overline{\bm{H}}_{n \perp} - \overline{\bm{E}}_{n \perp} \times
\bm{H}_{m\perp})dA & = & 0\nonumber \\
\label{EFpmA}
\end{eqnarray}
Therefore when $\zeta_m\neq\pm\overline{\zeta}$, by adding and subtracting the relations (\ref{EFOOA}) and (\ref{EFpmA}) we may deduce
\begin{equation}
\int_{\Omega} \bm{e}_z \cdot ( \bm{E}_{m\perp}\times
\overline{\bm{H}}_{n \perp} )dA= \int_{\Omega} \bm{e}_z \cdot ( \overline{\bm{E}}_{n\perp}\times
\bm{H}_{m \perp} )dA=0, \\ \label{EFOOB}
\end{equation}
which are the $ U_{mn}$ elements of the matrix $\bm{U}$ introduced earlier.  From (\ref{EFpmA}) and for $\zeta_m=\overline{\zeta}_n$ and real $\zeta_n$ we deduce that
the integral
\begin{equation}
\int_{\Omega} \bm{e}_z \cdot ( \bm{E}_{n\perp}\times
\overline{\bm{H}}_{n \perp} )dA  \\ \label{EFOOC}
\end{equation}
is real. When $\zeta_m$ is pure imaginary and
$\zeta_m=\zeta_n=i\zeta$ then from (\ref{EFpmA}) we deduce that
\begin{equation}
\int_{\Omega} \bm{e}_z \cdot ( \bm{E}_{n\perp}\times
\overline{\bm{H}}_{n \perp} )dA  \\ \label{EFOOD}
\end{equation}
is pure imaginary.

Now let us consider the case where $\zeta_n$ is complex. The $\bm{U}$ matrix is defined by
(\ref{AIE4}). For modes $\zeta_m=\zeta$ and $\zeta_n=-\overline{\zeta}$ from  (\ref{AIE4}) and (\ref{modeadj7}) we deduce
\begin{equation}
U_{mn}=\int_{\Omega} \bm{E}_{\zeta \perp}\cdot  (\bm{e}_z \times
\overline{\bm{H}}_{-\overline{\zeta} \perp})  dA =  -\int_{\Omega} \bm{E}_{\zeta \perp}\cdot  (\bm{e}_z \times
\bm{H}^\dag_{\zeta \perp})  dA.
\label{orto6}
\end{equation}
By using the orthonormal condition
\begin{equation}
\int_{\Omega} \bm{E}_{\zeta \perp}\cdot  (\bm{e}_z \times
\bm{H}^\dag_{\zeta \perp})  dA=1
\label{orto9}
\end{equation}
we find $U_{\zeta_m,\zeta_n}=U_{\zeta,-\overline{\zeta}}=-1$.

For the modes $\zeta_m=-\overline{\zeta}$ and $\zeta_n=\zeta$ from (\ref{AIE4}) we obtain
\begin{eqnarray}
U_{mn} & = &
\int_{\Omega}
\bm{E}_{-\overline{\zeta} \perp}\cdot  (\bm{e}_z \times
\overline{\bm{H}}_{\zeta \perp})dA \nonumber \\
& = & \int_{\Omega}
\overline{\bm{E}^\dag_{\zeta \perp}\cdot  (\bm{e}_z \times
\bm{H}_{\zeta \perp})}dA=1.
\label{orto7}
\end{eqnarray}
So the elements $U_{\zeta_m,\zeta_n}=U_{-\overline{\zeta},\zeta,} =
1$. This means that  $(\bm{U}+\bm{U}^H)/2$ is a diagonal matrix with
unit elements  on only the corresponding to propagating modal part.

%
%

%
%


\listoffigures


%
%

\end{document}